\documentclass[iop]{emulateapj}

\usepackage{amsmath}
\usepackage{textcomp}
\usepackage[colorlinks=true,citecolor=blue]{hyperref}
\usepackage{graphicx,color}
\usepackage{times}
\usepackage{amssymb}
\newcommand{\Msun}{\ensuremath{M_{\odot}}}
\newcommand{\phflux}{ph cm$^{-2}$ s$^{-1}$}
\newcommand{\lum}{erg\,s$^{-1}$}

\shorttitle{Radio-loud Narrow Line Seyfert 1 galaxies in $\gamma$-ray band}
\shortauthors{Paliya et al.}

\begin{document}

\title{{\it Fermi} Monitoring of Radio-Loud Narrow Line Seyfert 1 Galaxies}

\author{Vaidehi S. Paliya$^{1,\,2}$, C. S. Stalin$^{1}$, and C. D. Ravikumar$^{2}$} 
\affil{$^1$Indian Institute of Astrophysics, Block II, Koramangala, Bangalore-560034, India}
\affil{$^2$Department of Physics, University of Calicut, Malappuram-673635, India}
\email{vaidehi@iiap.res.in}

\begin{abstract}
We present detailed analysis of the $\gamma$-ray flux variability and spectral properties of the five radio-loud narrow line Seyfert 1 (RL-NLSy1) galaxies, detected by the Large Area Telescope onboard the {\it Fermi Gamma-ray Space Telescope}, namely 1H 0323+342, SBS 0846+513, PMN J0948+0022, PKS 1502+036, and PKS 2004$-$447. The first three sources show significant flux variations including the rapid variability of a few hours by 1H 0323+342. The average $\gamma$-ray spectrum of 1H 0323+342 and PMN J0948+0022 shows deviation from a simple power law (PL) behavior, whereas for other three sources, PL model gives better fit. The spectra of 1H 0323+342, SBS 0846+513, and PMN J0948+0022, respectively in low, flaring, and moderately active states, show significant curvature. Such curvature in the $\gamma$-ray spectrum of 1H 0323+342 and PMN J0948+0022 can be due to emission region located inside the broad line region (BLR) where the primary mechanism of the $\gamma$-ray emission is inverse-Compton (IC) scattering of BLR photons occurring in the Klein-Nishina regime. The $\gamma$-ray emission of SBS 0846+513 is explained by IC scattering of dusty torus photons which puts the emission region to be outside the BLR and thus under the Thomson regime. Therefore, the observed curvature of SBS 0846+513 could be intrinsic to the particle energy distribution. The presence of curvature in the $\gamma$-ray spectrum and flux variability amplitudes of some of the RL-NLSy1 galaxies suggest that these sources could be akin to low/moderate jet power flat spectrum radio quasars. 
\end{abstract}

\section{Introduction}\label{intro}
The launch of {\it Fermi Gamma-Ray Space Telescope} (hereafter {\it Fermi}) in 2008 has led to the discovery of $\gamma$-ray emission in a large number of blazars, comparable to that found by the {\it Energetic Gamma-Ray Experiment Telescope} \citep[{\it EGRET};][]{1993ApJS...86..629T}, owing to its large sensitivity and wide energy coverage. This has clearly demonstrated that blazars dominate the population of $\gamma$-ray emitters for which classification is known \citep{2012ApJS..199...31N}. Interestingly, {\it Fermi} has also observed variable $\gamma$-ray emission from some radio-loud narrow line Seyfert 1 (RL-NLSy1) galaxies \citep{2009ApJ...699..976A,2009ApJ...707L.142A,2012MNRAS.426..317D,2011MNRAS.413.2365C}. First identified as a new class of active galactic nuclei (AGN) by \citet{1985ApJ...297..166O}, NLSy1 galaxies have narrow Balmer lines (FWHM (H$_{\beta}$) $<$ 2000 km s$^{-1}$), weak [O~{\sc iii}] ([O~{\sc iii}]/H$_{\beta} <$ 3) and strong optical Fe~{\sc ii} lines \citep{1985ApJ...297..166O,1989ApJ...342..224G}. They also posses soft X-ray excess \citep{1996A&A...305...53B,1996A&A...309...81W,1999ApJS..125..297L}, relatively low black hole (BH) mass ($\sim$ 10$^{6}-10^{8}$ \Msun), high accretion rate \citep{2000ApJ...542..161P,2000NewAR..44..419H,2004ApJ...606L..41G,2006ApJS..166..128Z,2012AJ....143...83X}, and rapid X-ray flux variations \citep{1995MNRAS.277L...5P,1999ApJS..125..317L}. They also exhibit the radio-quiet/radio-loud dichotomy \citep{2000ApJ...543L.111L} and $\sim$ 7\% NLSy1 galaxies are found to be radio-loud \citep{2006AJ....132..531K}, a lesser fraction when compared to the $\sim$ 15\% found in quasar population \citep{1995PASP..107..803U}. Some of these RL-NLSy1 galaxies show compact core-jet structure, high brightness temperature and superluminal behavior \citep{2006AJ....132..531K,2006PASJ...58..829D}. Kilo-parsec scale radio structures are also reported in some RL-NLSy1 galaxies \citep{2008A&A...490..583A,2012ApJ...760...41D}. All these characteristics indicated the presence of aligned relativistic jets in RL-NLSy1 galaxies and the detection of $\gamma$-ray emission from five RL-NLSy1 galaxies by the Large Area Telescope (LAT) onboard {\it Fermi} provided the confirmation \citep{2009ApJ...699..976A,2009ApJ...707L.142A,2012MNRAS.426..317D}. Recently, high and variable optical polarization and rapid optical/infra-red flux variations are observed in some of these $\gamma$-ray emitting NLSy1 ($\gamma$-NLSy1) galaxies
\citep{2011PASJ...63..639I,2010ApJ...715L.113L,2013MNRAS.428.2450P,2012ApJ...759L..31J,2013ApJ...775L..26I}. Modeling the spectral energy distribution (SED) of $\gamma$-NLSy1 galaxies led to the conclusion that they posses the characteristic double hump SED similar to blazars and the physical properties of these sources are intermediate to the other two members of the blazar family, namely BL Lac objects and 
flat spectrum radio quasars (FSRQs) \citep{2009ApJ...707L.142A,2012MNRAS.426..317D,2011MNRAS.413.1671F,2013ApJ...768...52P}. Also, \citet{2010ASPC..427..243F} have shown that one of these $\gamma$-NLSy1 galaxies, PMN J0948+0022 occupy an intermediate position in the well known blazar sequence \citep{1998MNRAS.299..433F}. Though, many properties of $\gamma$-NLSy1 galaxies are similar to that of blazars, there exist characteristic differences as well. For example, blazars are known to be hosted by massive elliptical galaxies \citep{2009arXiv0909.2576M}, whereas, the $\gamma$-NLSy1 galaxies, similar to other NLSy1 galaxies, are likely to be hosted by spiral galaxies with relatively low-mass BHs at their centers. Therefore, detailed studies of these $\gamma$-NLSy1 galaxies will enable us understand the physical properties of relativistic jets at different mass and accretion rate scales. Also, a clear understanding of the shape of the observed $\gamma$-ray spectrum can help us place constraints on the various models proposed to explain the high energy component present in broadband SED of these sources.

In this work, we present $\gamma$-ray flux variability properties and $\gamma$-ray spectral properties of five $\gamma$-NLSy1 galaxies using $\sim$ 5 years of {\it Fermi}-LAT data. Wherever possible, we also study their $\gamma$-ray spectral behavior in different activity states. The motivation here is to see the similarities and/or differences in both the $\gamma$-ray spectra as well as the $\gamma$-ray flux variability of these sources with respect to the blazar class of AGN that emits copiously in the $\gamma$-ray band.

The paper has been organized as follows. Sample selection is discussed in Section~\ref{sample}. Section~\ref{data_red} is devoted to data reduction and analysis procedures. We discuss the results in Section~\ref{result} followed by our conclusions in Section~\ref{conclusion}. Throughout the work, we adopt $\Omega_{m}$ = 0.27, $\Omega_{\varLambda}$ = 0.73 and {\it H}$_{0}$ = 71 km s$^{-1}$ Mpc$^{-1}$.

\section{Sample}\label{sample}
As of now (2014 September 01), five RL-NLSy1 galaxies are found to be emitting in the $\gamma$-ray band by {\it Fermi}-LAT with high significance, having test statistic (TS) $>$ 25 \citep[$\sim$ 5$\sigma$;][]{1996ApJ...461..396M}. General information about these five $\gamma$-NLSy1 galaxies is given in Table~\ref{general}. In Table~\ref{general}, the values of right ascension, declination, redshift, and the apparent V-band magnitude are taken from \citet{2010A&A...518A..10V}. The given radio spectral indices are calculated using the 6 cm and 20 cm flux densities found in 
\citet{2010A&A...518A..10V} ($S_{\nu} \propto \nu^{\alpha}$) and the values of radio loudness parameter R\footnote{defined as the ratio of flux density in 5 GHz to that in optical {\it B}-band} are taken from \citet{2011nlsg.confE..24F}.

\section{Data reduction and Analysis}\label{data_red}
The observational data used in this work were collected by {\it Fermi}-LAT \citep{2009ApJ...697.1071A} over the first five years of its operation, from 2008 August 05 (MJD 54,683) to 2013 November 01 (MJD 56,597). For the analysis of the data, standard procedures described in the {\it Fermi}-LAT documentation are used. We use {\tt ScienceTools v9r32p5} along with the post-launch instrument response functions (IRFs) P7REP\_SOURCE\_V15. In the energy range of 0.1$-$300 GeV, only events belonging to the SOURCE class are selected. To avoid contamination from Earth limb $\gamma$-rays, source events coming from zenith angle $<$ 100$^{\circ}$ are extracted. A filter of ``\texttt{DATA$\_$QUAL==1}'',  ``\texttt{LAT$\_$CONFIG==1}'', and ``\texttt{ABS(ROCK\_ANGLE)$<$52}'' is used to select the good time intervals. To model the background, galactic diffuse emission component (gll\_iem\_v05.fits) and an isotropic component  iso\_source\_v05.txt\footnote{http://fermi.gsfc.nasa.gov/ssc/data/access/lat/BackgroundModels.html} are considered. While fitting the five year average spectrum, the normalization of components in the background model as well as the photon index of the galactic diffusion model are allowed to vary freely. We fix these parameters to the values obtained from the fitting during the subsequent time series analysis.

\begin{figure*}
\vbox
 {
\hbox{
\hspace{-1cm}
      \includegraphics[width=10cm,height=7.0cm]{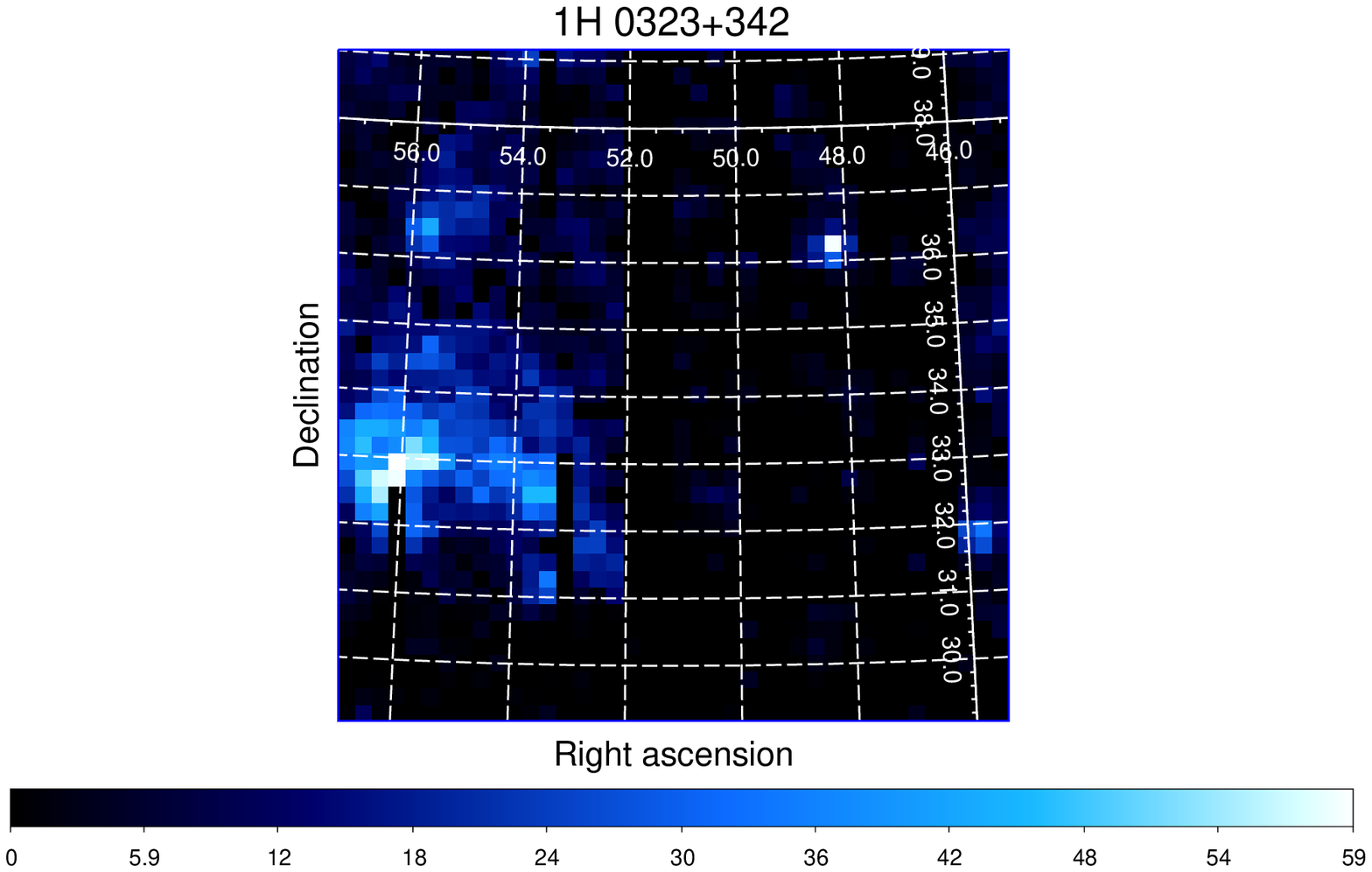}
      \includegraphics[width=10cm,height=7.0cm]{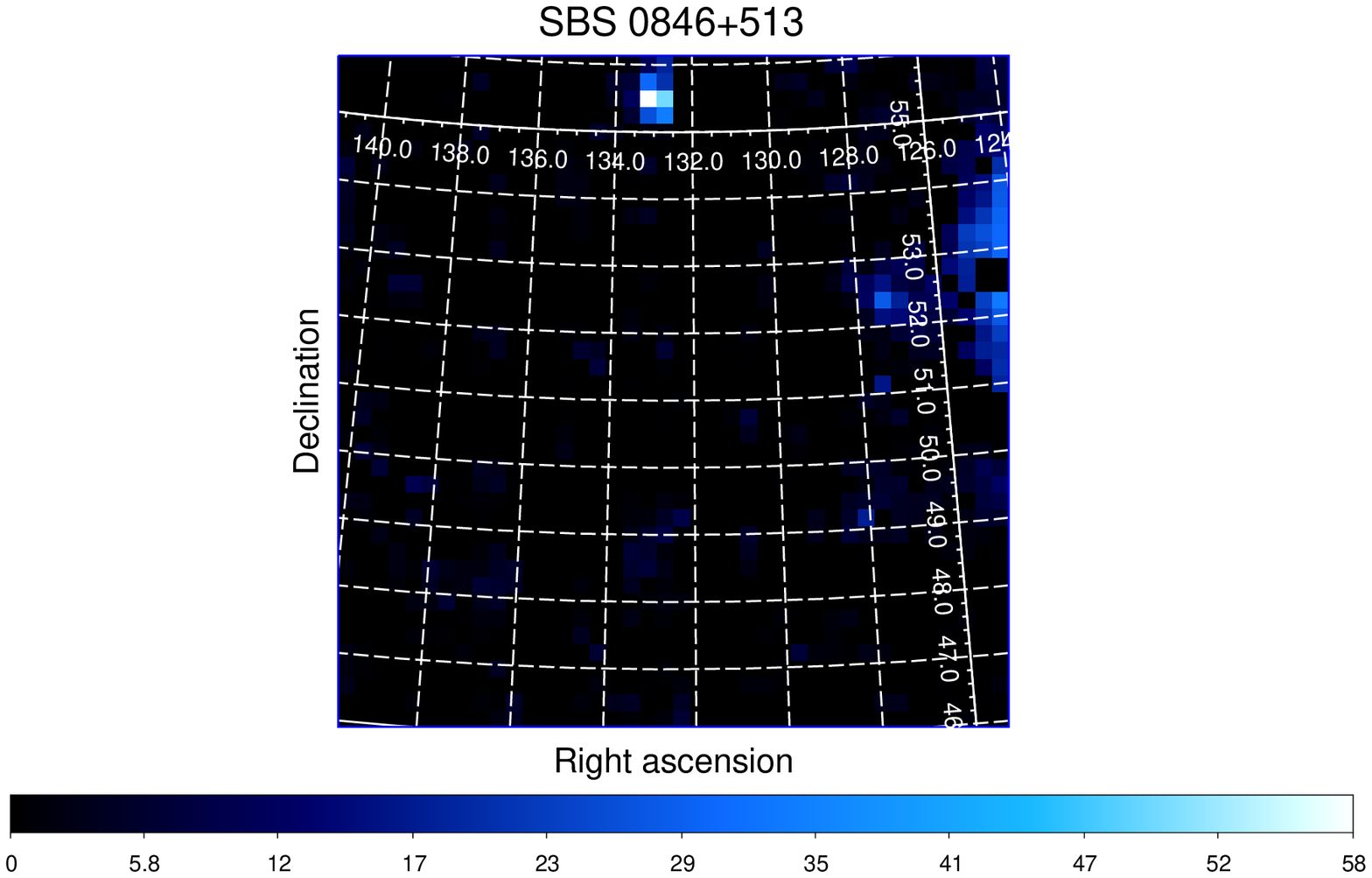}
     }
\hbox{\hspace{-1cm}
      \includegraphics[width=10cm,height=7.0cm]{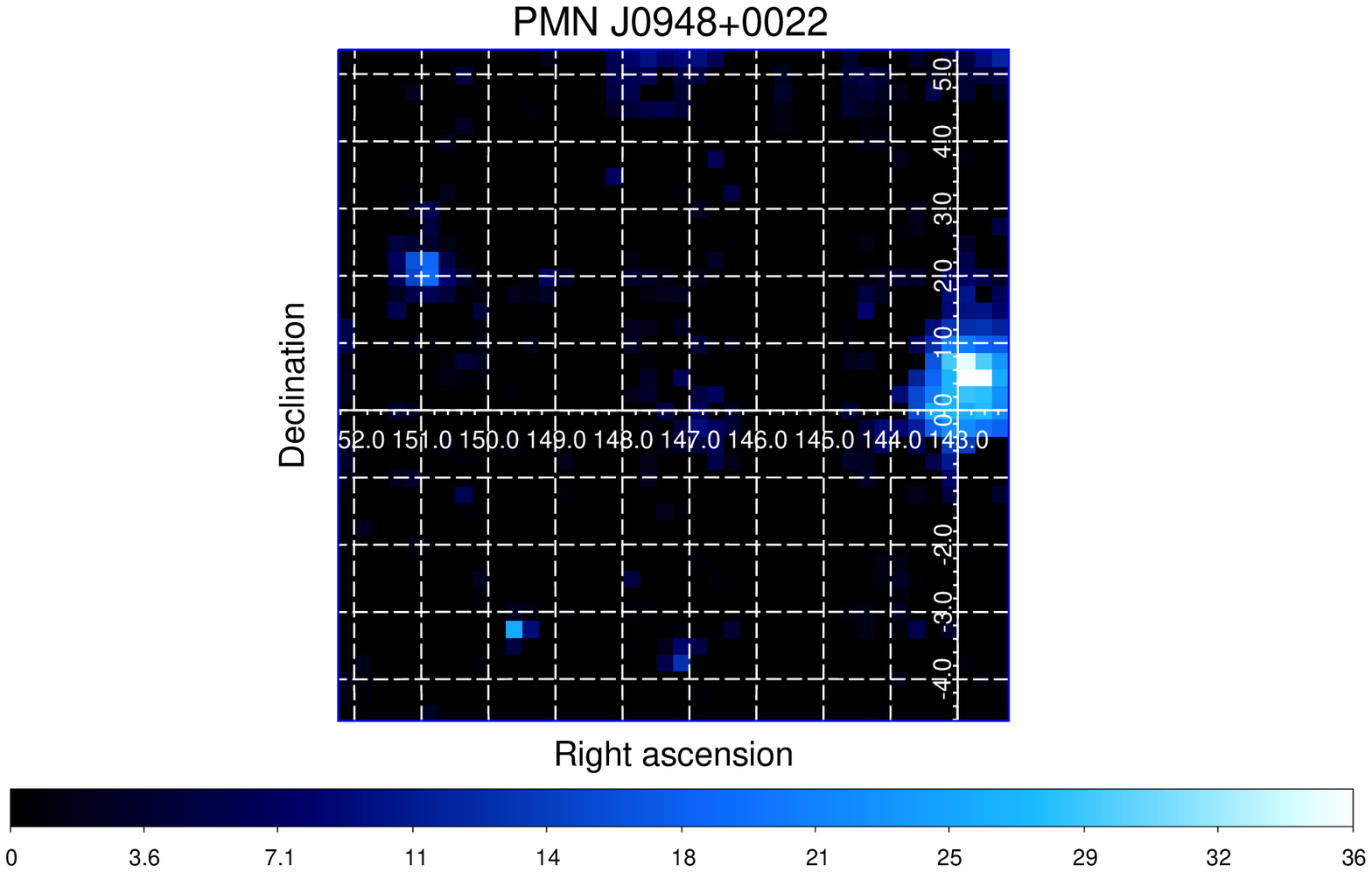}
      \includegraphics[width=10cm,height=7.0cm]{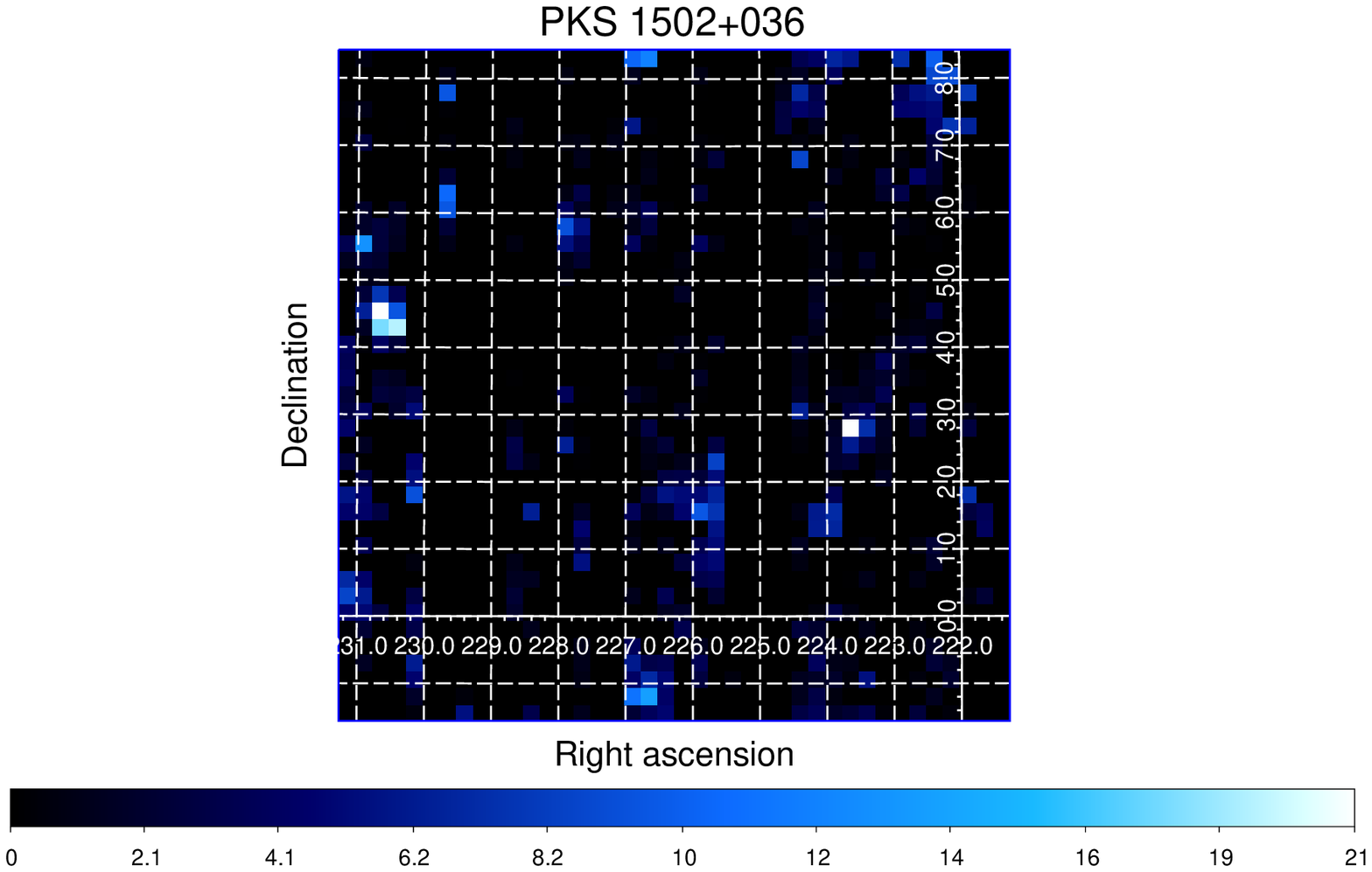}
     } \centering
      \includegraphics[width=10cm,height=7.0cm]{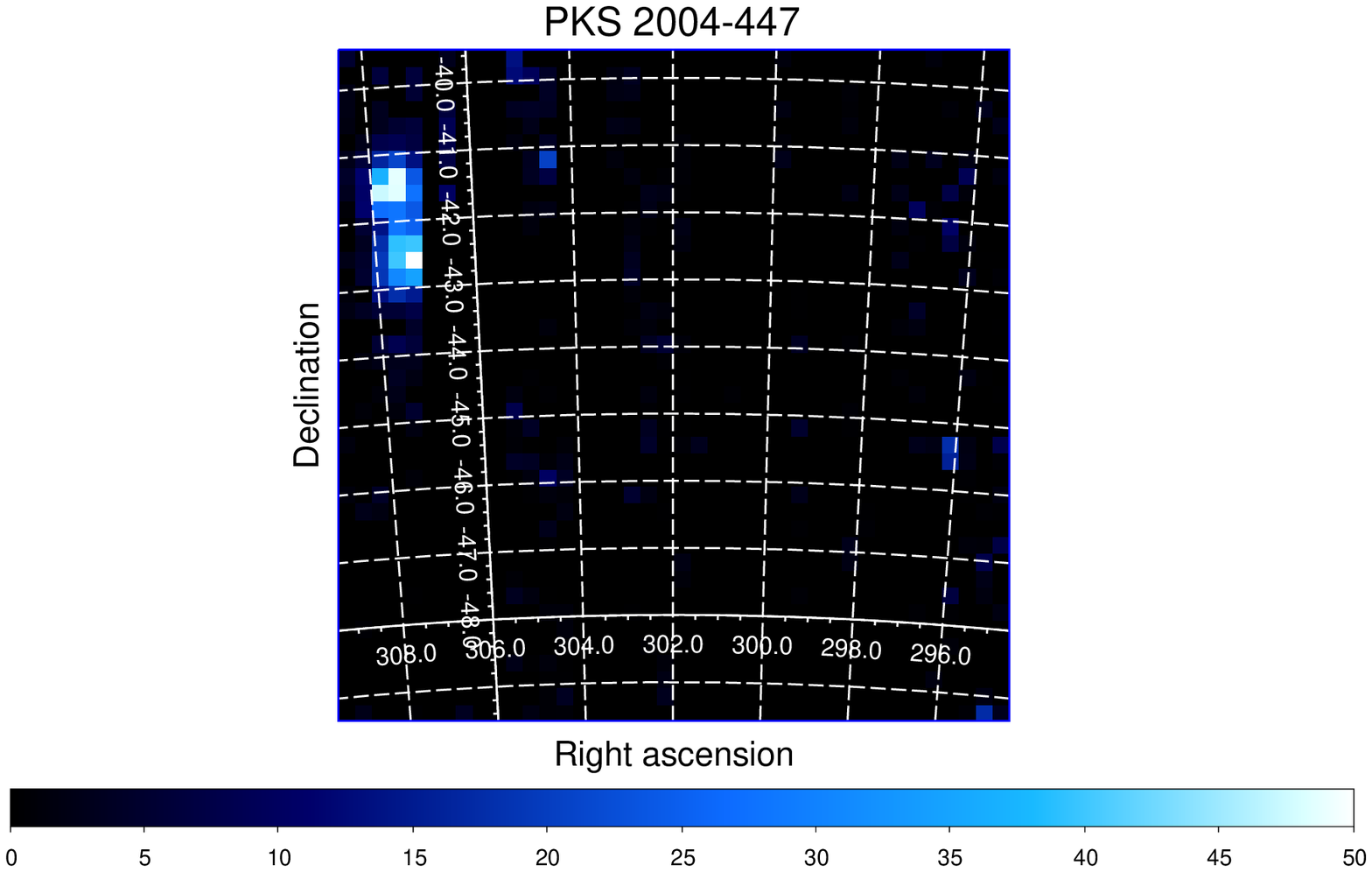}
}
\caption{The residual TS maps of the 0.1$-$300 GeV events during the time period covered in this work, centered on the co-ordinates of the $\gamma$-NLSy1 galaxies. There are two new sources each in the TS maps of 1H 0323+342 and PKS 2004$-$447 while one new source each in the TS maps of SBS 0846+513 and PMN J0948+0022. We do not find any new source in the TS map of PKS 1502+036.}\label{fig:TSMAPS}
\end{figure*}

To evaluate the significance of the $\gamma$-ray signal we use the maximum-likelihood (ML) test statistic TS = 2$\Delta$ log(likelihood) between models with and without a point source at the position of the source of interest. We use the binned likelihood method to generate the average $\gamma$-ray spectrum as well as the spectra over different time periods. The source model includes all the point sources from the second {\it Fermi}-LAT catalog \citep[2FGL;][]{2012ApJS..199...31N} that lies within the 15$^{\circ}$ region of interest (ROI) of every source. Many sources are detected by {\it Fermi}-LAT after the release of the 2FGL catalog \citep[see e.g.][]{2011ATel.3579....1G} and thus not included in it. However, if these newly detected sources lie in the vicinity of a source of interest, they could have an effect on the results of the analysis. We therefore, search for possible new sources within the ROI of the five $\gamma$-NLSy1 galaxies studied here, by generating their residual TS map, using the tool {\tt gttsmap}. These residual TS maps are shown in Figure~\ref{fig:TSMAPS}. Using the criterion TS $> 25$, we find two un-modeled sources associated with each of the residual TS maps of 1H 0323+342 (RA, Dec. = 48$^{\circ}$.236, 36$^{\circ}$.260 and 56$^{\circ}$.092, 32$^{\circ}$.804; J2000) and PKS 2004$-$447 (RA, Dec. = 307$^{\circ}$.257, $-$42$^{\circ}$.592 and 307$^{\circ}$.505, $-$41$^{\circ}$.542; J2000) and one source each in the residual TS maps of SBS 0846+513 (RA, Dec. = 133$^{\circ}$.112, 55$^{\circ}$.497; J2000) and PMN J0948+0022 (RA, Dec. = 142$^{\circ}$.854, 0$^{\circ}$.497; J2000). No new source with TS $> 25$ is seen in the residual TS map of PKS 1502+036. These new sources are then modeled by a power law (PL) and are included in our analysis. However, we note that inclusion of these new sources have negligible effect on the results obtained without including them as they are faint. Further, both PL and log parabola (LP) models are used to parameterize the $\gamma$-ray spectra of the sources. For objects that fall within 7$^{\circ}$ of each of the target sources studied here, all parameters except the scaling factor are allowed to vary, for the sources that lie between 7$^{\circ}$ to 14$^{\circ}$ only the normalization factor is allowed to vary freely and for the remaining sources that lie outside 14$^{\circ}$, all parameters are fixed to the values given in the 2FGL catalog. We perform a first run of the ML analysis and then remove all sources from the model with TS $<$ 25. This updated model is then used to carry out a second ML analysis. If the likelihood fitting shows non-convergence, all parameters of the sources lying outside 7$^{\circ}$ from the center of the ROI are fixed to the values obtained from the average likelihood fitting and the fitting is repeated. In cases where we find the fitting to show non-convergence, we repeat the fit by fixing the photon indices of the sources further 1$^{\circ}$ inside from the edge of the ROI and this process is carried out iteratively till the analysis converges. To generate weekly binned light curves, we use a PL model for all $\gamma$-NLSy1 galaxies, as the PL indices obtained from this model show smaller statistical uncertainties when compared to those obtained from fits using more complex models. To generate daily averaged as well as six hour binned light curves, only the normalization parameter of the source of interest is kept free and the rest of the parameters are fixed to the values obtained from the five years averaged fitting. In all the cases, i.e. both for light curve as well as spectrum generation, we consider the source to be detected at the $\sim$ 3$\sigma$ level \citep[e.g.][]{1991ApJ...374..344K} if the condition TS $>$ 9 is satisfied in the respective time or energy bin. We calculate 2$\sigma$ upper limits whenever $\bigtriangleup~F_{\gamma}/F_{\gamma} > 0.5$ and/or 1 $<$ TS $<$ 9, where $\bigtriangleup~F_{\gamma}$ is the 1$\sigma$ error estimate in the $\gamma$-ray flux $F_{\gamma}$. This is found by varying the flux of the source given by {\tt gtlike}, till the TS reaches a value of 4 \citep[e.g.][]{2010ApJS..188..405A}. The upper limits are not calculated for TS $<$ 1. Primarily governed by the uncertainty in the effective area, the measured fluxes have energy dependent systematic uncertainties of around 10\% below 100 MeV, 5\% between 316 MeV to 10 GeV and 10\% above 10 GeV\footnote{http://fermi.gsfc.nasa.gov/ssc/data/analysis/LAT\_caveats.html}. Unless otherwise specified, all the errors quoted in this paper are 1$\sigma$ statistical uncertainties.

\section {Results and Discussion}\label{result}
\subsection{Long Term Flux Variability}\label{long_var}
\begin{figure*}
\centering
\includegraphics[width=18cm,height=12cm]{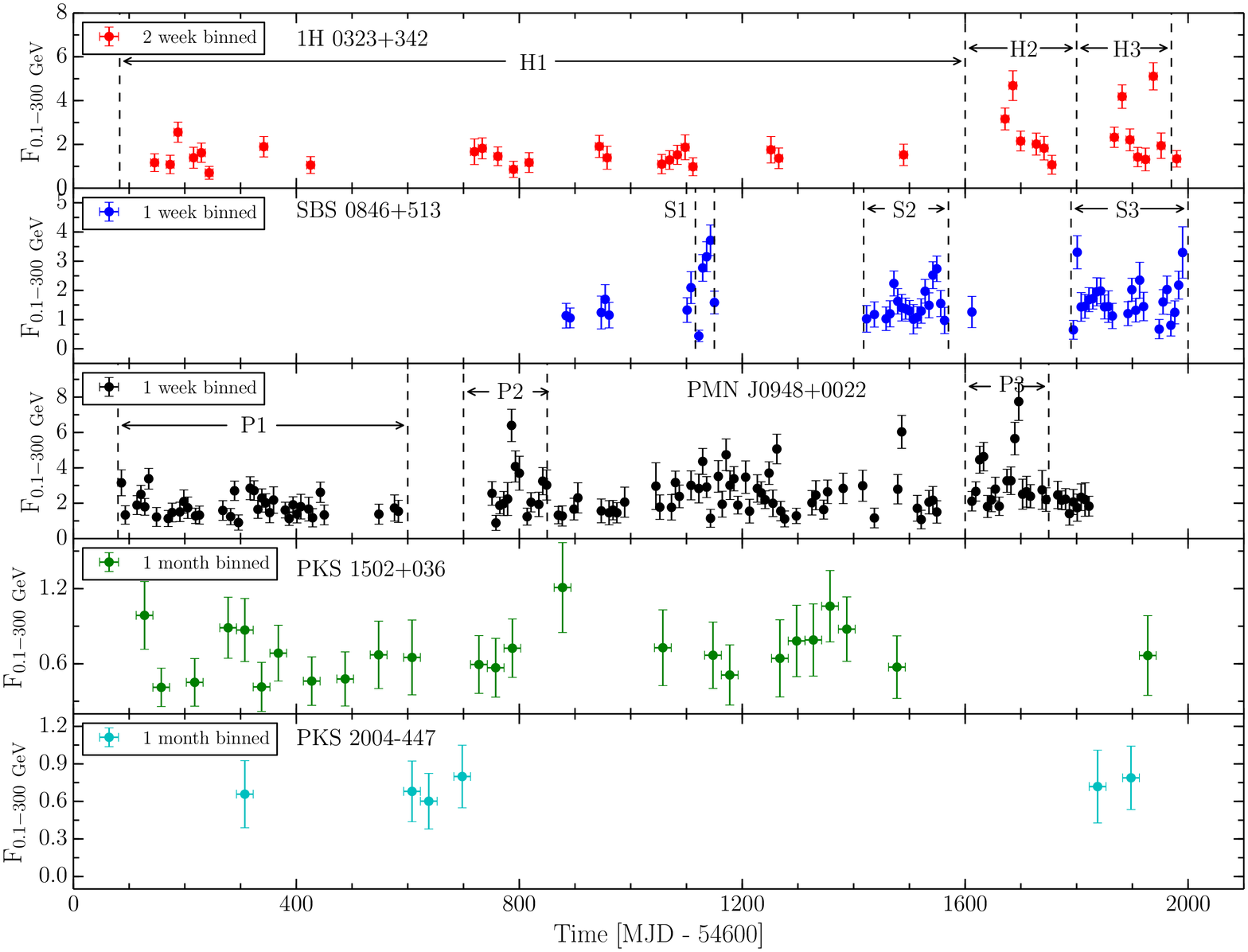}
\caption{Long term $\gamma$-ray light curves of the $\gamma$-NLSy1 galaxies. Flux values are in units of 10$^{-7}$ \phflux. MJD 54,600 corresponds to 2008 May 14. Different activity periods are marked with appropriate notations. See text for details.}\label{lc}
\end{figure*}

The $\gamma$-ray light curves of the five $\gamma$-NLSy1 galaxies, obtained using the procedures outlined in Section~\ref{data_red}, are shown in Figure~\ref{lc}. Depending on the $\gamma$-ray brightness, different time bins are used for different objects. For clarity, we have not shown upper limits. Variability in the light curve is tested by calculating fractional rms variability ($F_{\rm var}$) and variability probability by means of a simple $\chi^{2}$ test, using the recipes of \citet{2003MNRAS.345.1271V} and \citet{2010ApJ...722..520A} respectively and the results are shown in Table~\ref{statistic}.

Bi-weekly binned light curve of 1H 0323+342 shows that it is nearly steady in $\gamma$-rays during the initial four years of the LAT observations, however, displays flaring behavior at two epochs in 2013, one around MJD 56,300 and the other around MJD 56,500. They are marked as H2 and H3 in Figure~\ref{lc}. The $F_{\rm var}$ for this source is found to be 0.498~$\pm$~0.046, which is on the higher side of that obtained for FSRQs ($\sim 0.24 \pm 0.01$) by \citet{2010ApJ...722..520A}, thus indicating substantial flux variations. The probability that the source has shown variation is $>$~99\% (see Table~\ref{statistic}). The average $\gamma$-ray flux during most of its quiescence is found to be $\sim$ 8 $\times$ 10$^{-8}$ \phflux. A maximum $\gamma$-ray flux of (5.11 $\pm$ 0.62) $\times$ 10$^{-7}$ \phflux~is observed in the bin centered at MJD 56,537 when a GeV flare was detected by {\it Fermi}-LAT \citep{2013ATel.5344....1C,2014ApJ...789..143P}. Five years averaged TS value of this source is found to be $\sim$ 723 which corresponds to $\sim$ 27$\sigma$ detection.

SBS 0846+513 was in quiescence during the first two years of {\it Fermi} operation and thus was not included in the 2FGL catalog. This source was discovered in the $\gamma$-ray band only when {\it Fermi}-LAT detected a GeV flare during 2011 June \citep{2011ATel.3452....1D}. After that, many flaring activities have been observed from this source and they are shown by S1, S2, and S3 in Figure~\ref{lc}. A flux of $\sim$ 4 $\times$ 10$^{-8}$ \phflux~and TS $\sim$ 1700 is obtained from the average analysis. The probability that this source is variable is $>$~99\% and shows a value of 0.467~$\pm$~0.042 to the $F_{\rm var}$.

PMN J0948+0022 is the first RL-NLSy1 galaxy detected in the $\gamma$-ray band by {\it Fermi}-LAT. An average analysis of $\sim$ 5 years of the LAT data gives the $\gamma$-ray flux and TS of $\sim$ 1.3 $\times$ 10$^{-7}$ \phflux~and $\sim$ 3300 respectively. From its weekly binned light curve, the peak flux value of (7.74~$\pm$~1.07)~$\times$~10$^{-7}$ \phflux~is observed during a recent GeV flare \citep{2013ATel.4694....1D}. The ratio of the maximum to minimum flux is obtained as $\sim$17 which is significantly higher than that noticed by \citet{2012A&A...548A.106F}. The primary reason of this difference is due to the time ranges covered by both the analyses. The peak flux is observed in the bin centered at MJD 56,297, which is not covered by \citet{2012A&A...548A.106F}. Few significant flaring activities are observed and we consider two of them for further analysis. These flaring periods are noted as P2 and P3 in Figure~\ref{lc}. We find a value of $F_{\rm var}$ of 0.443~$\pm$~0.029. This clearly hints at the existence of substantial flux variability in this source.

PKS 1502+036 has been detected by {\it Fermi}-LAT most of the times, however, at a low flux level of $\sim$ 4.5 $\times$ 10$^{-8}$ \phflux, as can be seen in its monthly binned light curve in Figure~\ref{lc}. The five years averaged analysis gives a TS value of $\sim$ 435 which corresponds to $\sim$~20$\sigma$ detection. On comparing the results obtained here with that of \citet{2013MNRAS.433..952D}, the average flux and photon index values are found to be quite similar despite the fact that we have included one more year of the LAT data as compared to theirs. The similarity seen in both analyses can be justified by the fact that PKS 1502+036 is hardly detected by {\it Fermi}-LAT in the period MJD 56,235$-$56,597 (see the third panel from top in Figure~\ref{lc}) which is not covered by them. The $\chi^2$ analysis classifies the source to be a non-variable.

PKS 2004$-$447 has the lowest values for the $\gamma$-ray flux (1.42~$\pm$~0.24~$\times$~10$^{-8}$~\phflux) and the TS (106) among $\gamma$-NLSy1 galaxies and was detected by {\it Fermi}-LAT only on a few occasions. Hence no definite conclusion can be drawn about its flux variability behavior owing to the low photon statistics as we could not get a significant value for the $F_{\rm var}$.

\subsection{Short Term Flux Variability and Jet Energetics}\label{small_var}
Blazars are known to vary on sub-hour time scales \citep[e.g.,][]{2011A&A...530A..77F,2013ApJ...766L..11S} and this could constrain the size as well as the location of the $\gamma$-ray emitting region \citep[see e.g.,][]{2010MNRAS.405L..94T,2013MNRAS.431..824B}. We, therefore, search for the presence of flux variability over short time scales (of the order of hours) in the light curves of the $\gamma$-NLSy1 galaxies. From the long term light curves of 1H 0323+342, SBS 0846+513, and PMN J0948+0022, we identify the epochs of high activity (shown in Figure~\ref{lc}) and generate one day as well as six hour binned light curves. We then scan the data to calculate the shortest flux doubling/halving time scale using the following equation:
\begin{equation*}
F(t)=F(t_{0}).2^{(t-t_{0})/\tau}
\end{equation*}
where $F(t)$ and $F(t_0)$ are the fluxes at time $t$ and $t_0$, respectively, and $\tau$ is the characteristic doubling/halving time scale. We also set the condition that the difference in flux at the epochs $t$ and $t_0$ is at least significant at 3$\sigma$ level \citep{2011A&A...530A..77F}.

\begin{figure*}
\centering
\includegraphics[width=12cm,height=6cm]{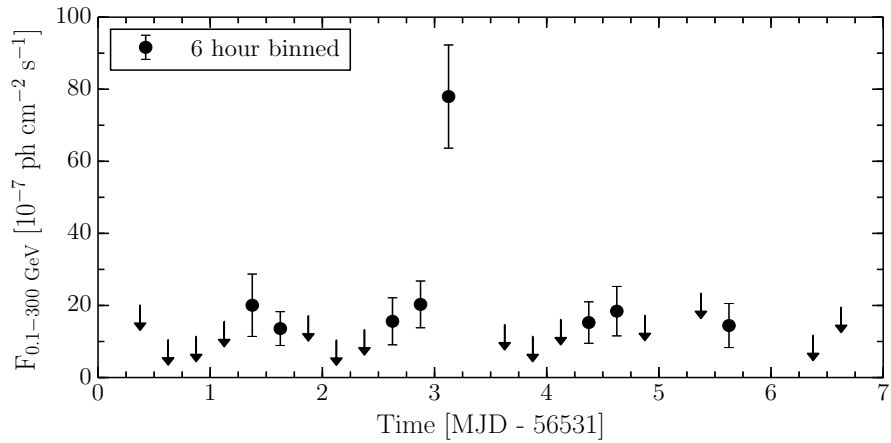}
\caption{Six hour binned $\gamma$-ray light curve of 1H 0323+342 covering the period of GeV outburst. Upper limits at 2$\sigma$ level are shown by downward arrows.}\label{3hr_6hr}
\end{figure*}

The recent GeV flare detected from 1H 0323+342 \citep{2013ATel.5344....1C} had its daily $\gamma$-ray flux as high as (1.75 $\pm$ 0.33) $\times$ 10$^{-6}$ \phflux~which is $\sim$~23 times larger compared to the five years averaged flux. The photon statistics during this epoch (2013 August 28 to 2013 September 1) is good enough to generate six hour binned light curve (see Figure~\ref{3hr_6hr}). In contrast to the earlier observations where $\gamma$-ray variability of the $\gamma$-NLSy1 galaxies were characterized over longer time scales \citep[$\geqslant$ 1 day; e.g.,][]{2011MNRAS.413.2365C,2012A&A...548A.106F}, this epoch of high activity provides evidence of extremely fast variability observed from a $\gamma$-NLSy1 galaxy. A flux doubling time as small as 3.09~$\pm$~0.85 hours (assuming exponential rise of the flare) is the fastest $\gamma$-ray variability ever observed from this class of AGN. In the time bin when the highest flux is measured, the TS value is 136 associated with a registered count of 34. The flux doubling time obtained here is significantly different from the value of 3.3~$\pm$~2.5 days reported by \citet{2011MNRAS.413.2365C} using the method of e-folding time scale. The primary reason for this contrast is the fact that 1H 0323+342 was in quiescence for most of the first four years of {\it Fermi} operation and hence could not produce enough photon statistics for studying short time scale variability. These findings are similar to that recently reported by \citet{2014ApJ...789..143P}. The highest flux in the six hour binned light curve is measured as (7.80 $\pm$ 1.43) $\times$ 10$^{-6}$ \phflux, which corresponds to an isotropic $\gamma$-ray luminosity of 2.82 $\times$ 10$^{46}$ \lum, which is $\sim$~100 times higher than the five years averaged value. Total power radiated in the $\gamma$-ray  band \citep{1997ApJ...484..108S}, thus, would be $L_{\gamma,em} \simeq L_{\gamma,iso}/2\Gamma^2 \simeq$ 2.20 $\times$ 10$^{44}$ \lum~\citep[assuming a bulk Lorentz factor $\Gamma$=8;][]{2014ApJ...789..143P}, which is approximately 16\% of the Eddington luminosity considering a black hole mass of 10$^{7}$ \Msun~\citep{2006ApJS..166..128Z}.

Detection of a large flare in 2011 June led to the discovery of SBS 0846+513 in the $\gamma$-ray band. During this period of high activity, we find a maximum daily averaged flux of (8.20 $\pm$ 0.08) $\times$ 10$^{-7}$ \phflux, which is almost 20 times higher than its five years averaged flux. When the data are further re-binned using six hour bins, we notice a high $\gamma$-ray flux of (9.37 $\pm$ 1.74) $\times$ 10$^{-7}$ \phflux. The apparent isotropic $\gamma$-ray luminosity is found to be $\sim$ 1.44 $\times$ 10$^{48}$ \lum. Blazars, in general, and FSRQs in particular, are known to be emitters of such high power in the $\gamma$-ray band. If a bulk Lorentz factor of 15 \citep{2012MNRAS.426..317D} is assumed, we estimate that the total radiated power in $\gamma$-rays would be $L_{\gamma,em}~\simeq$ 3.21 $\times$ 10$^{45}$ \lum~which is about 93\% of the Eddington luminosity assuming a BH mass of 10$^{7.4}$ \Msun~\citep{2008ApJ...685..801Y}, a value generally found for powerful FSRQs \citep[e.g.][]{2012Sci...338.1445N}. We find many instances during this flaring period when the apparent isotropic $\gamma$-ray luminosity exceeds 10$^{48}$ \lum~(e.g., MJD 55,742 and 55,745). During the GeV flare in 2011 June, we find the shortest flux halving time scale of $\sim$ 25.6~$\pm$~11.0 hours. 

PMN J0948+0022 is the most $\gamma$-ray luminous NLSy1 galaxy ever detected by {\it Fermi}-LAT. During its recent GeV flare \citep{2013ATel.4694....1D} we observe a maximum one day averaged flux of (1.76 $\pm$ 0.32) $\times$ 10$^{-6}$ \phflux~on MJD 56,293 which is about 14 times the five years averaged value and the highest daily binned flux observed from the source in our analysis. This highest one day binned flux is similar to that noted by \citet{2012A&A...548A.106F}. On further dividing this active period into six hour bins, we find a maximum flux value of (2.40 $\pm$ 0.83) $\times$ 10$^{-6}$ \phflux. This corresponds to an isotropic $\gamma$-ray luminosity of 1.87 $\times$ 10$^{48}$ \lum~which is $\sim$ 20 times its five years averaged value. Total radiated power in the $\gamma$-ray band is $L_{\gamma,em}~\simeq 4.16 \times 10^{45}$ \lum~(assuming a bulk Lorentz factor of $\Gamma = 15$). If its BH mass is assumed to be $10^{7.5} M_{\odot}$~\citep{2008ApJ...685..801Y}, we find that the radiated $\gamma$-ray power is $\sim$ 95\% of the Eddington luminosity. The shortest time scale of variability observed from this source is found to be 74.7~$\pm$~27.6 hours, which is comparable to that reported by \citet{2012A&A...548A.106F} and \citet{2011MNRAS.413.2365C}.

\subsection{Gamma-ray Spectrum}\label{gamma_spec}
Analysis of the $\gamma$-ray spectral shape is done using two spectral models: PL ({$dN/dE \propto E^{-\Gamma_{\gamma}}$}), where $\Gamma_{\gamma}$ is the photon index and LP ({ $dN/dE \propto (E/E_{o})^{-\alpha-\beta log({\it E/E_{o}})}$}) where $E_o$ is the reference energy which is 
fixed at 300 MeV, $\alpha$ is the photon index at $E_o$ and $\beta$ is the curvature index which defines the curvature around the peak. The low photon statistics of all the sources precludes us to analyze the $\gamma$-ray spectral shape using more complex models such as broken power law (BPL) or power law with exponential cut-off. The spectral analysis is performed using a binned ML estimator {\tt gtlike} following the prescription given in Section~\ref{data_red}. We use a likelihood ratio test \citep{1996ApJ...461..396M} to check the PL model (null hypothesis) against the LP model (alternative hypothesis). Following \citet{2012ApJS..199...31N}, the curvature of test statistic $TS_{\rm curve}$ = 2(log L$_{\rm LP}-$log L$_{\rm PL}$) is also calculated. Presence of a significant curvature is tested by setting the condition $TS_{\rm curve} > 16$.

\subsubsection{Average Spectral Analysis}\label{avg_spec}
\begin{figure*}
\centering
\vbox
 {
\hbox{
      \includegraphics[width=9cm,height=7cm]{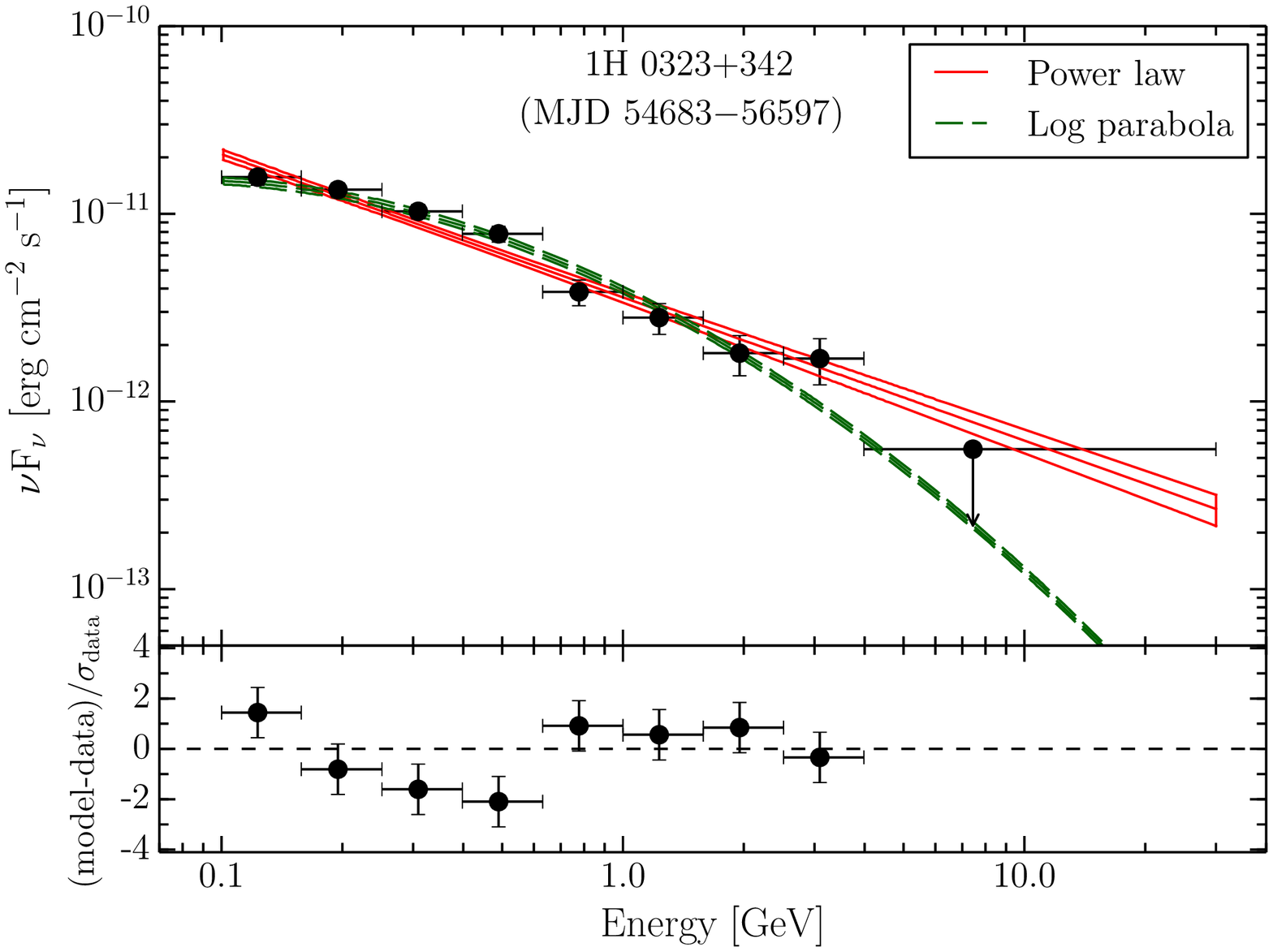}
      \includegraphics[width=9cm,height=7cm]{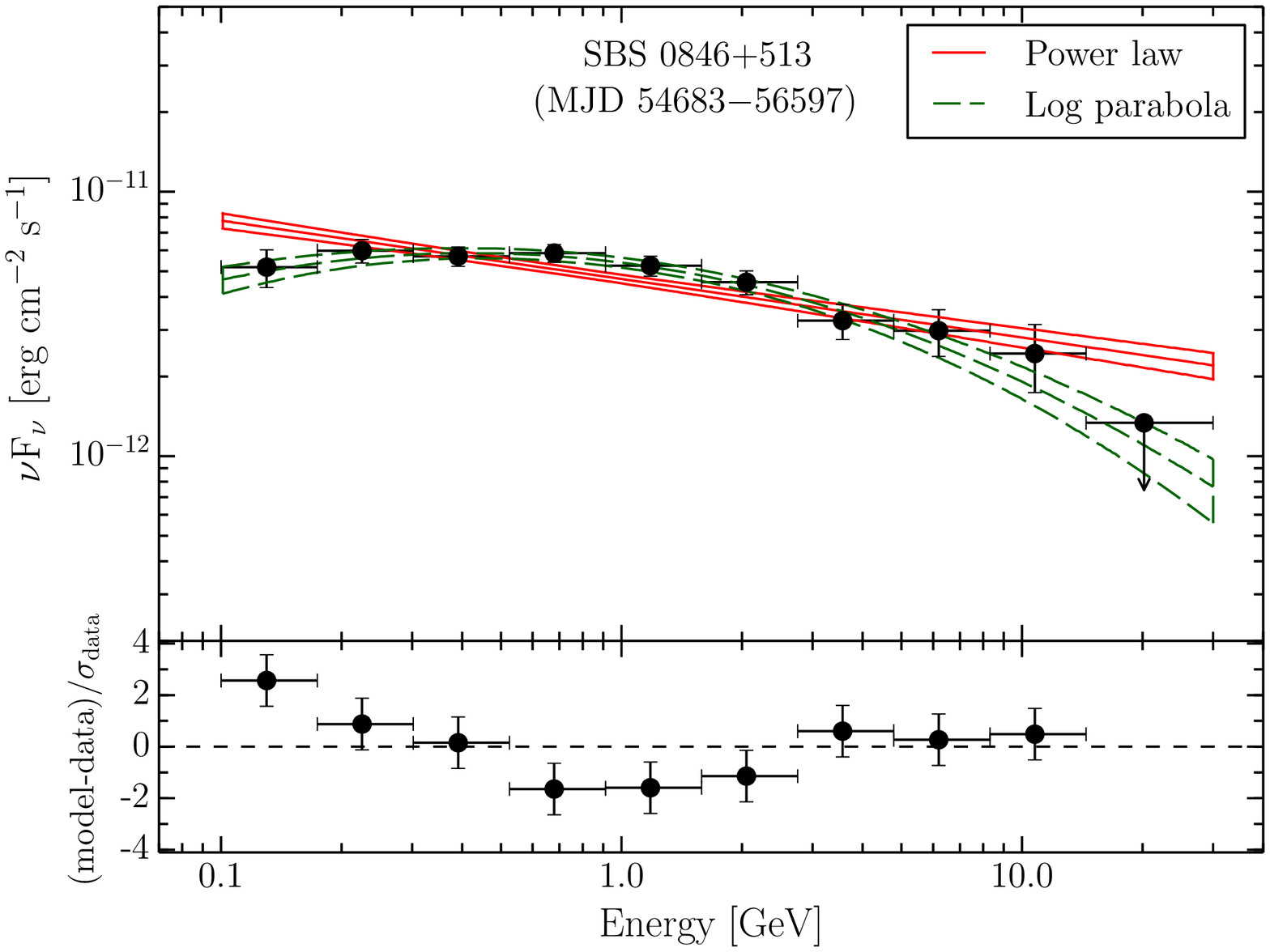}
     }
\hbox{
      \includegraphics[width=9cm,height=7cm]{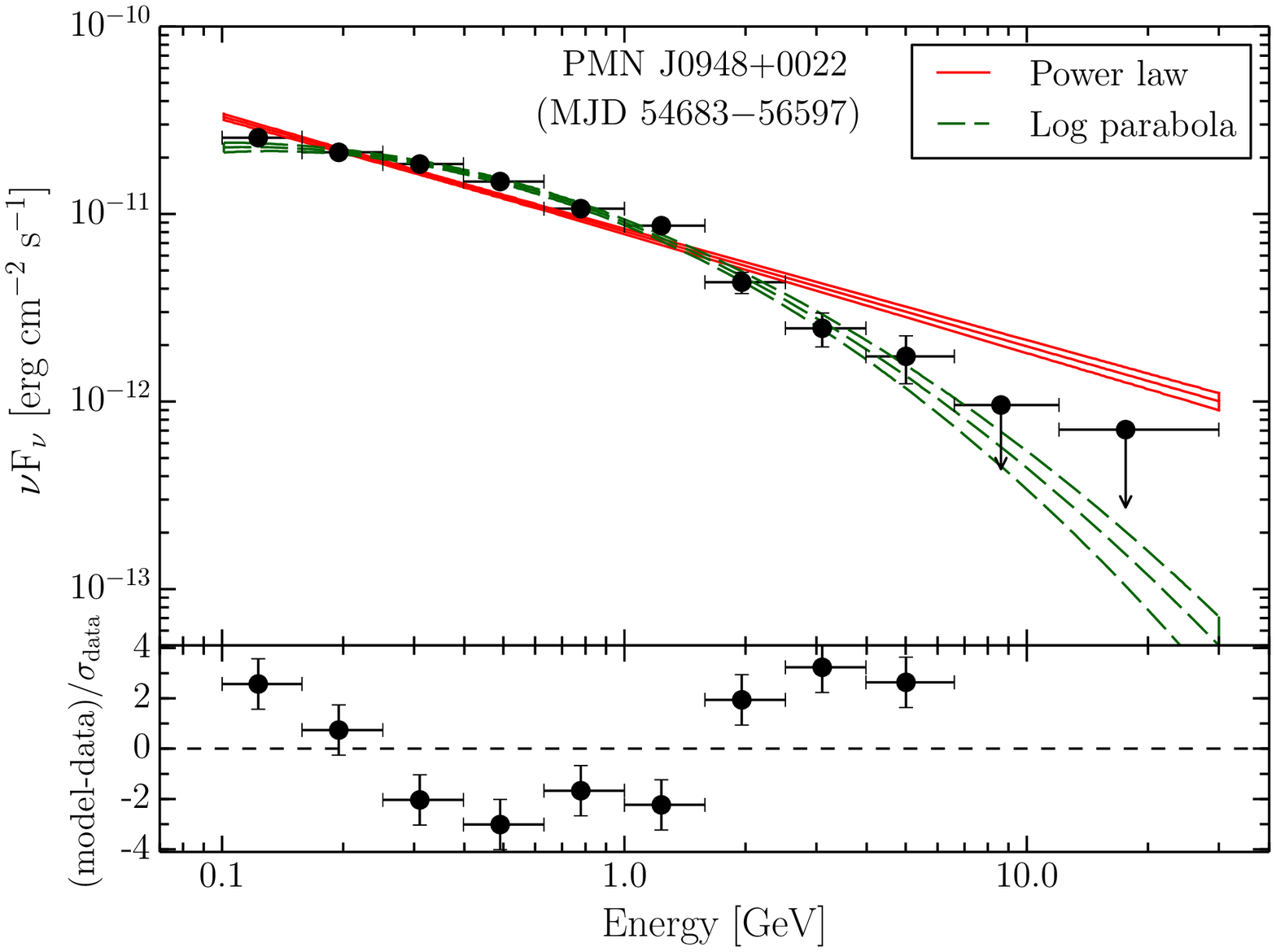}
      \includegraphics[width=9cm,height=7cm]{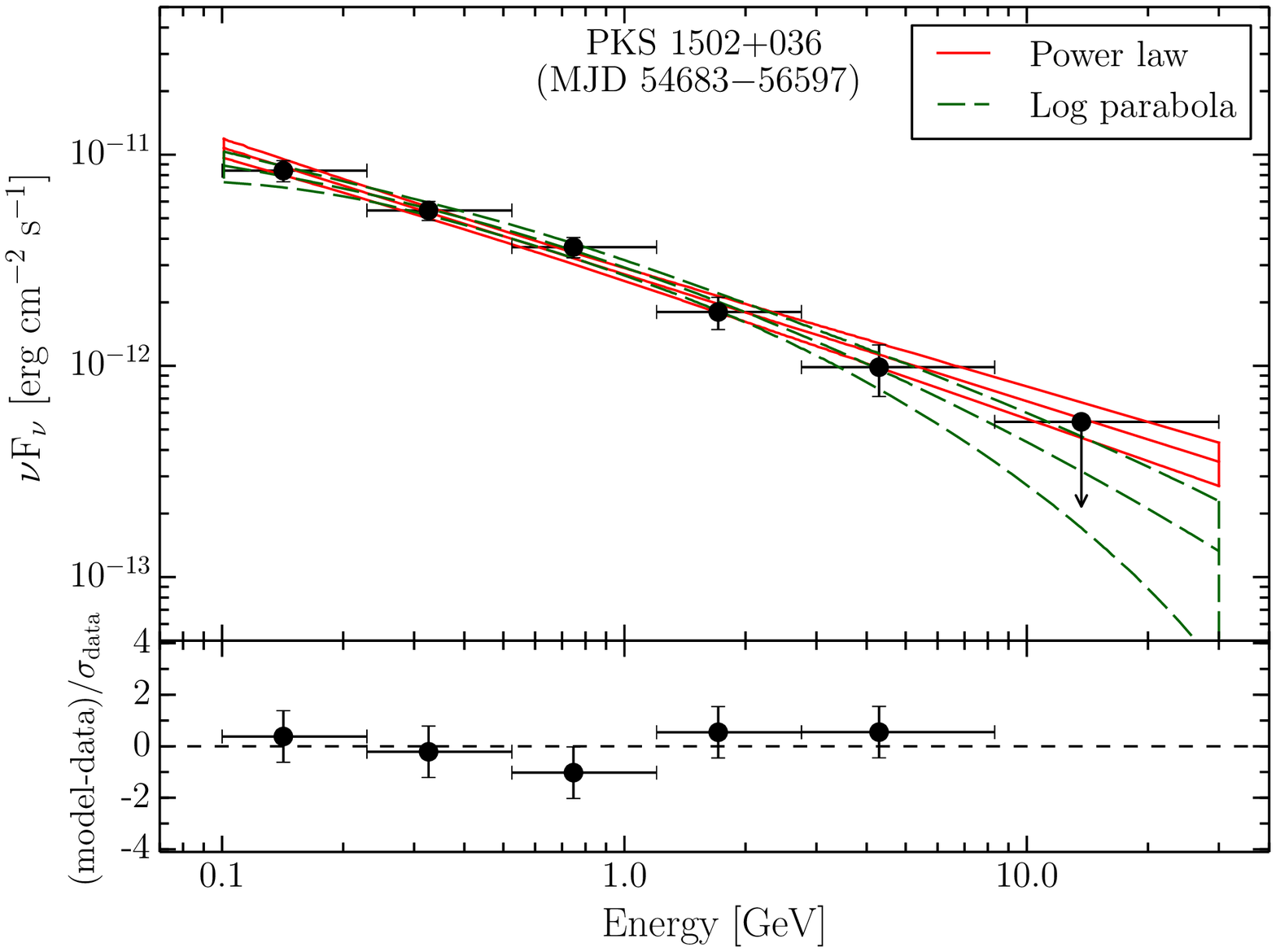}
     }
\includegraphics[width=9cm,height=7cm]{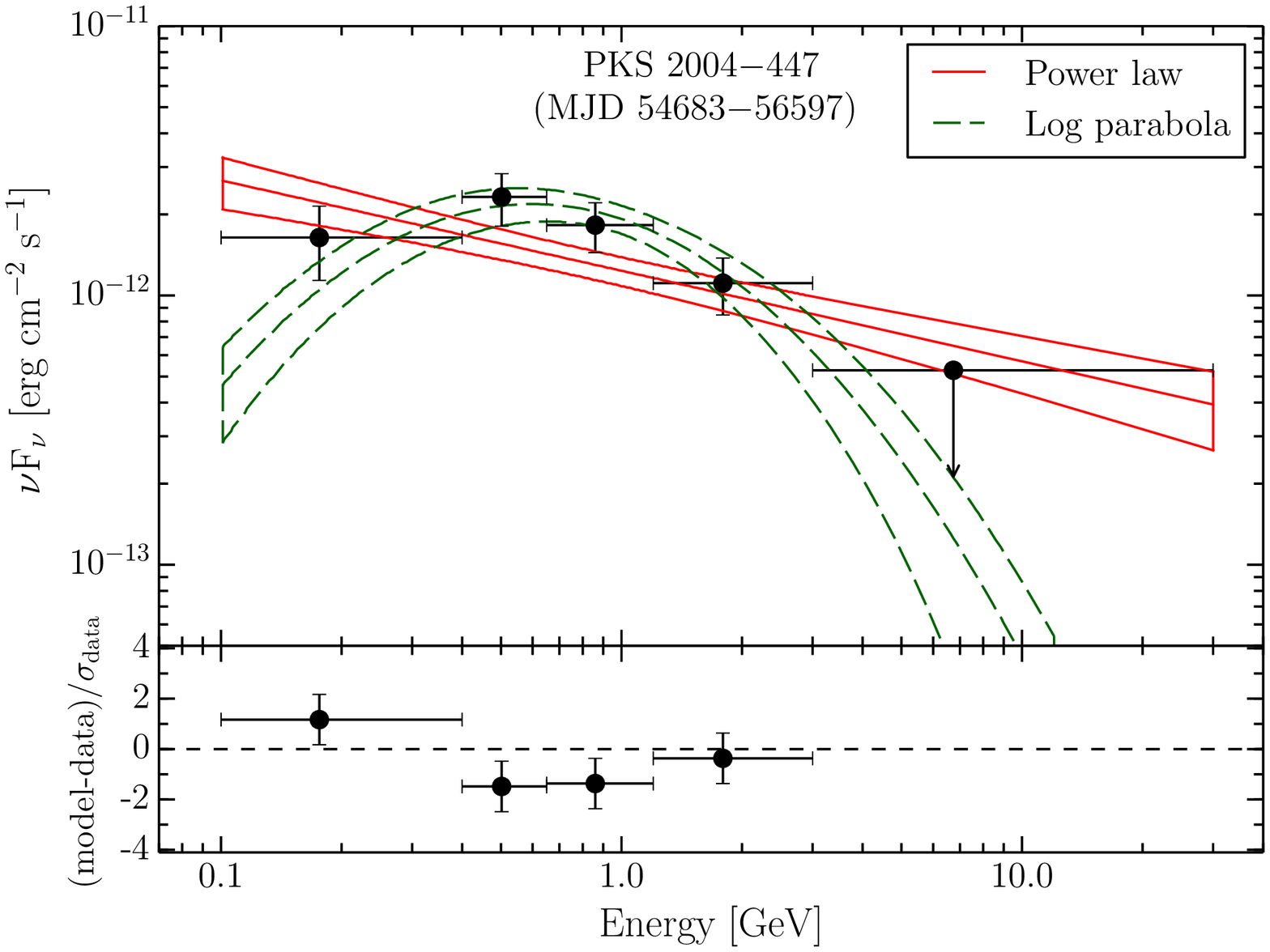}
  }
\caption{Five years averaged {\it Fermi}-LAT SEDs of the $\gamma$-NLSy1 galaxies. Solid and dashed lines represent PL and LP model fits respectively with uncertainties. The residuals in the lower panels refer to the PL model.}\label{avg_spec_fig}
\end{figure*}

The results of the average spectral analysis of the five $\gamma$-NLSy1 galaxies using five years of {\it Fermi}-LAT data are tabulated in Table~\ref{avg_res}. Two sources, 1H 0323+342 and PMN J0948+0022, are found to have a significant curvature in their $\gamma$-ray spectrum. The value of $TS_{\rm curve}$ reported in Table~\ref{avg_res} shows that the LP model describes their spectrum better than the PL model. A $TS_{\rm curve}$ value of $\sim$~15 for SBS 0846+513 indicates a possible curvature in its $\gamma$-ray spectrum. For the remaining two sources, no significant curvature is noticed. This is possibly due to low photon statistics. However, in future, when more $\gamma$-ray data would be accumulated, it is likely that they too show a significant deviation from the PL model. To show the departure from a PL behavior and for better visualization of the spectral shape, we separately perform a binned likelihood analysis on appropriately chosen energy bins covering 0.1$-$300 GeV and generate the SEDs. The results are shown in Figure~\ref{avg_spec_fig}. Since there is no signal beyond 30 GeV, the $\gamma$-ray spectra are shown up to 30 GeV only. In this figure, the solid red line represents the PL model while the LP model is plotted by green dashed line. 
From the residual plot (lower panel) of the PL model fits to the sources, it is evident that the spectrum of 1H 0323+342 and PMN J0948+0022 clearly deviate from the PL behavior, confirming the $TS_{\rm curve}$ test. Also, hints of curvature can be seen in the residual plot of SBS 0846+513. PMN J0948+0022 has the most prominent curvature in its $\gamma$-ray spectrum ($\sim 8\sigma$) whereas $\sim 4\sigma$ significant curvature is noticed in the spectrum of 1H 0323+342. Further, the curvature index $\beta$ for PKS 2004$-$447 is found to be high and appears non-realistic as there is no significant curvature found in the $\gamma$-ray spectrum of this source. Thus, the unusual $\beta$ value may be indicating the poor photon statistics associated with the observation and hence seems unreliable.

\subsubsection{Spectral Analysis of Low and High Activities}\label{flr_spec}
\begin{figure*}
\hbox{
\includegraphics[width=9.0cm,height=7.0cm]{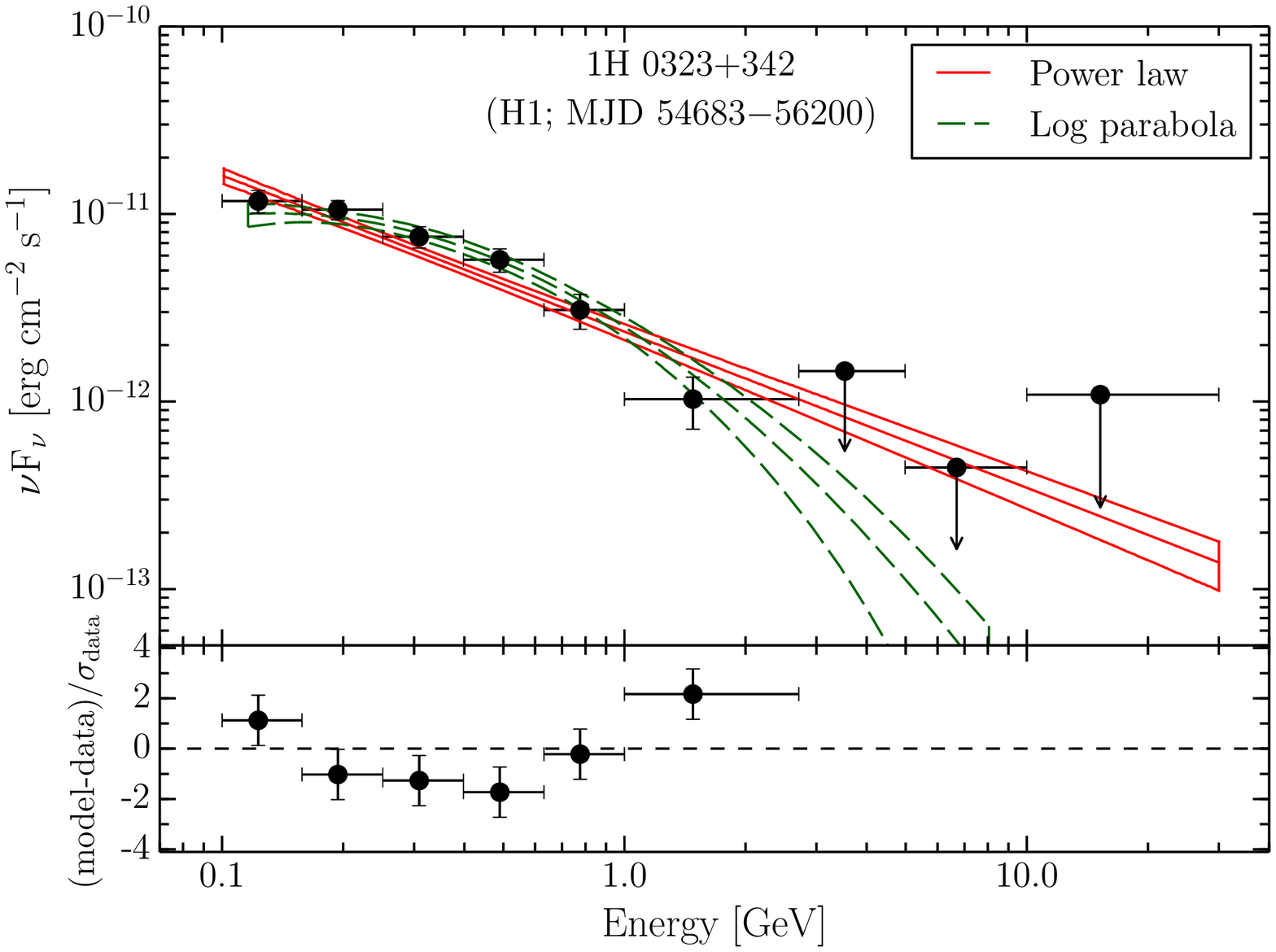}
\includegraphics[width=9.0cm,height=7.0cm]{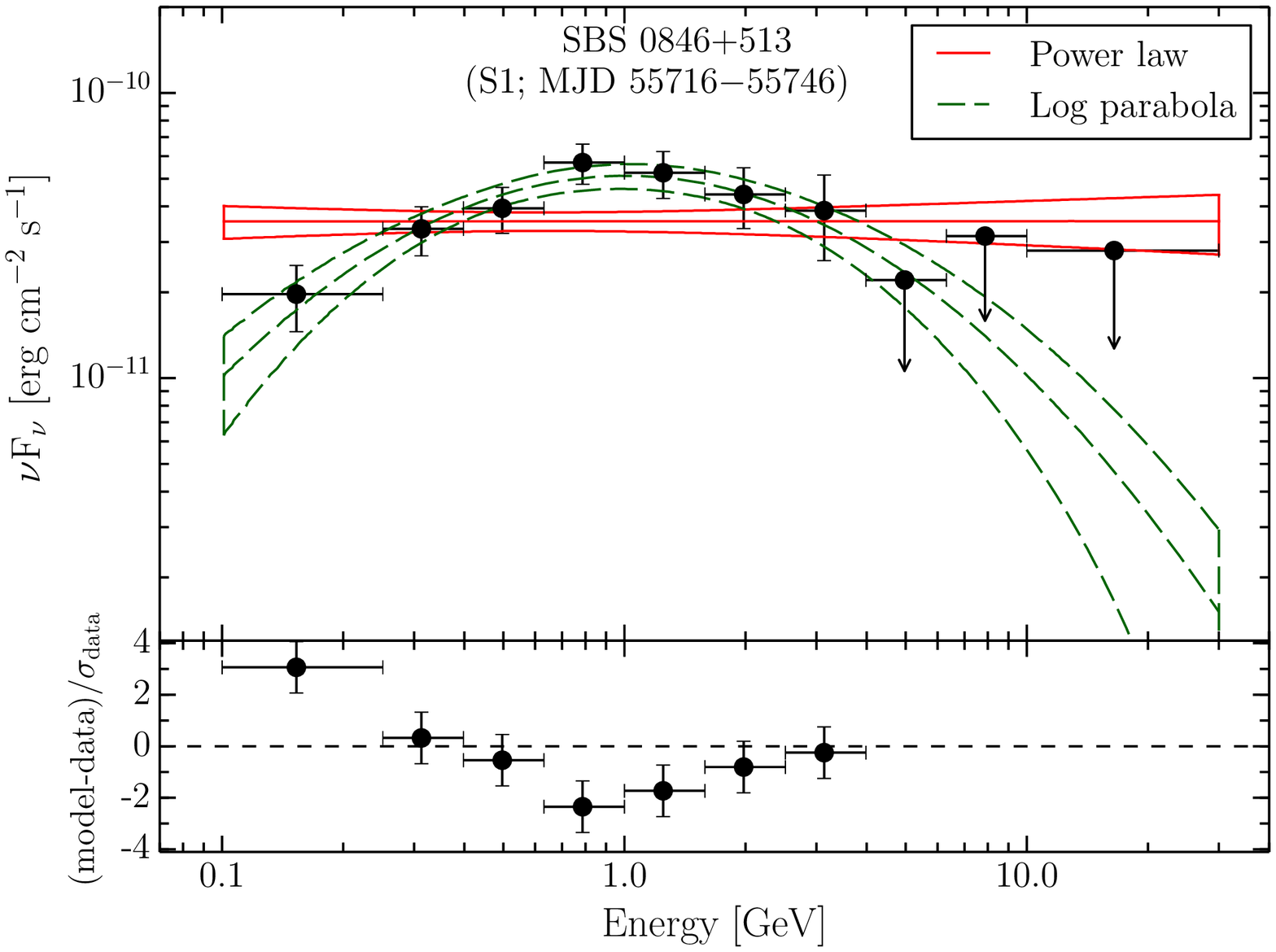}
}
\hbox{
\includegraphics[width=9.0cm,height=7.0cm]{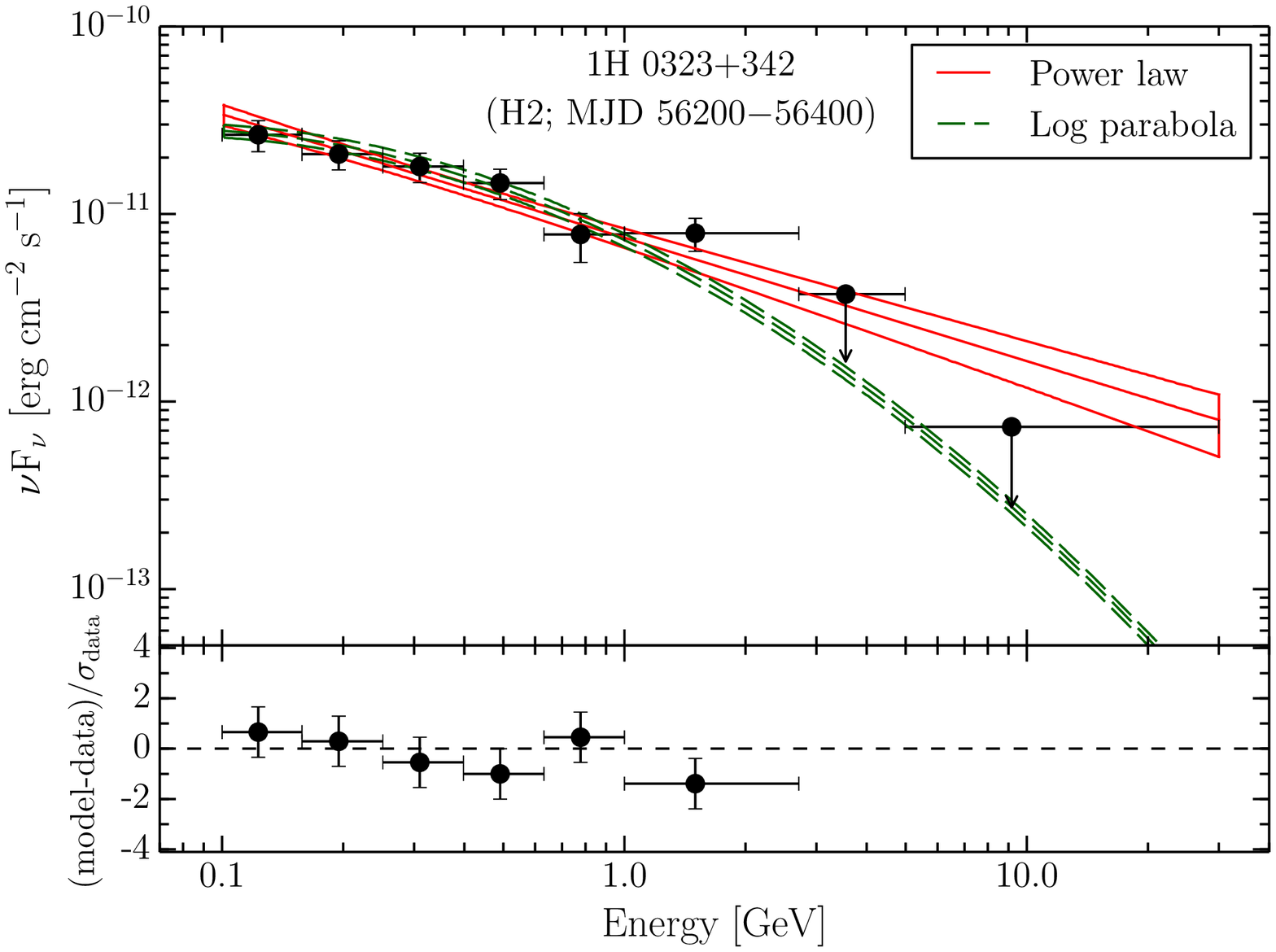}
\includegraphics[width=9.0cm,height=7.0cm]{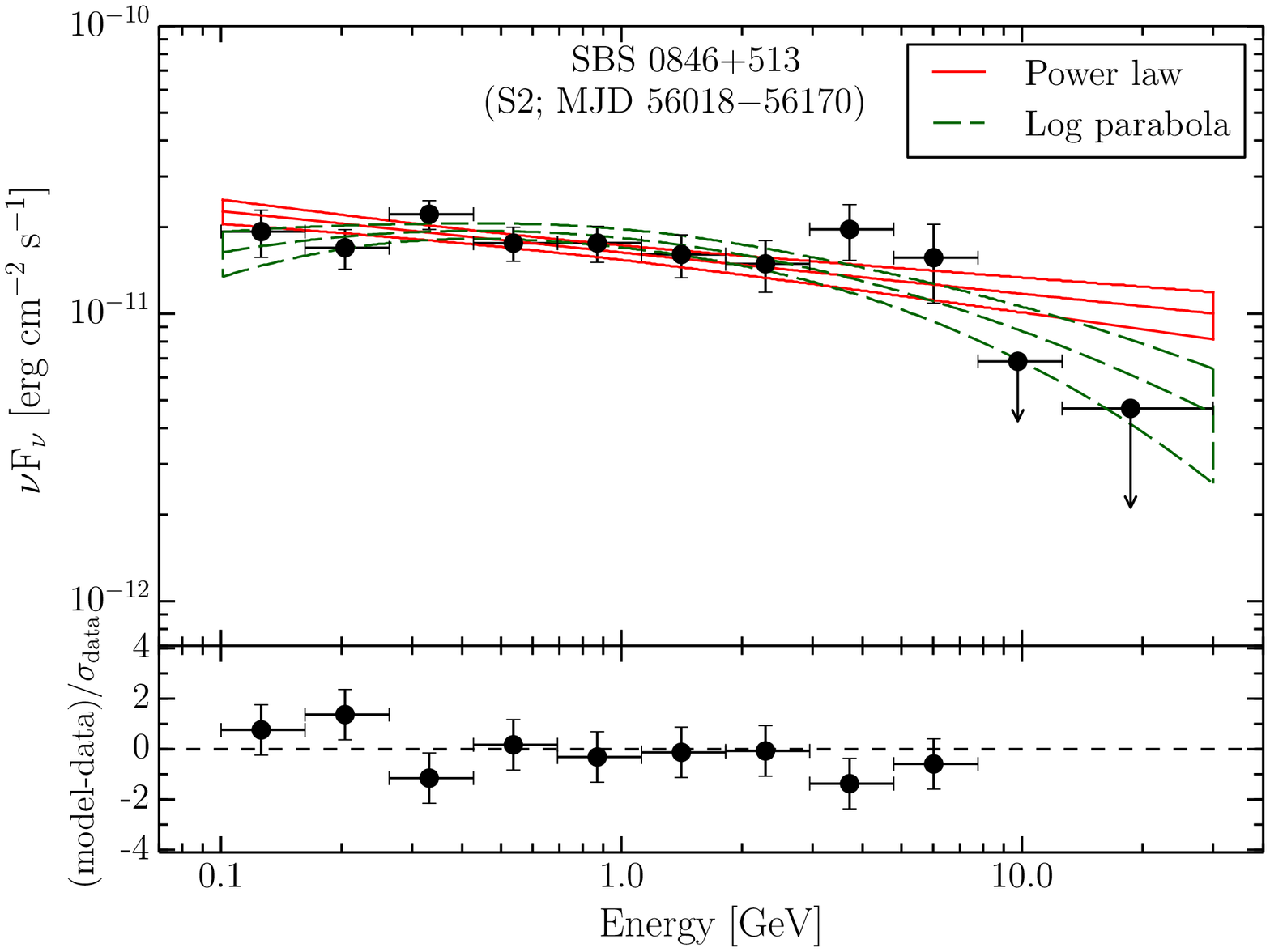}
}
\hbox{
\includegraphics[width=9.0cm,height=7.0cm]{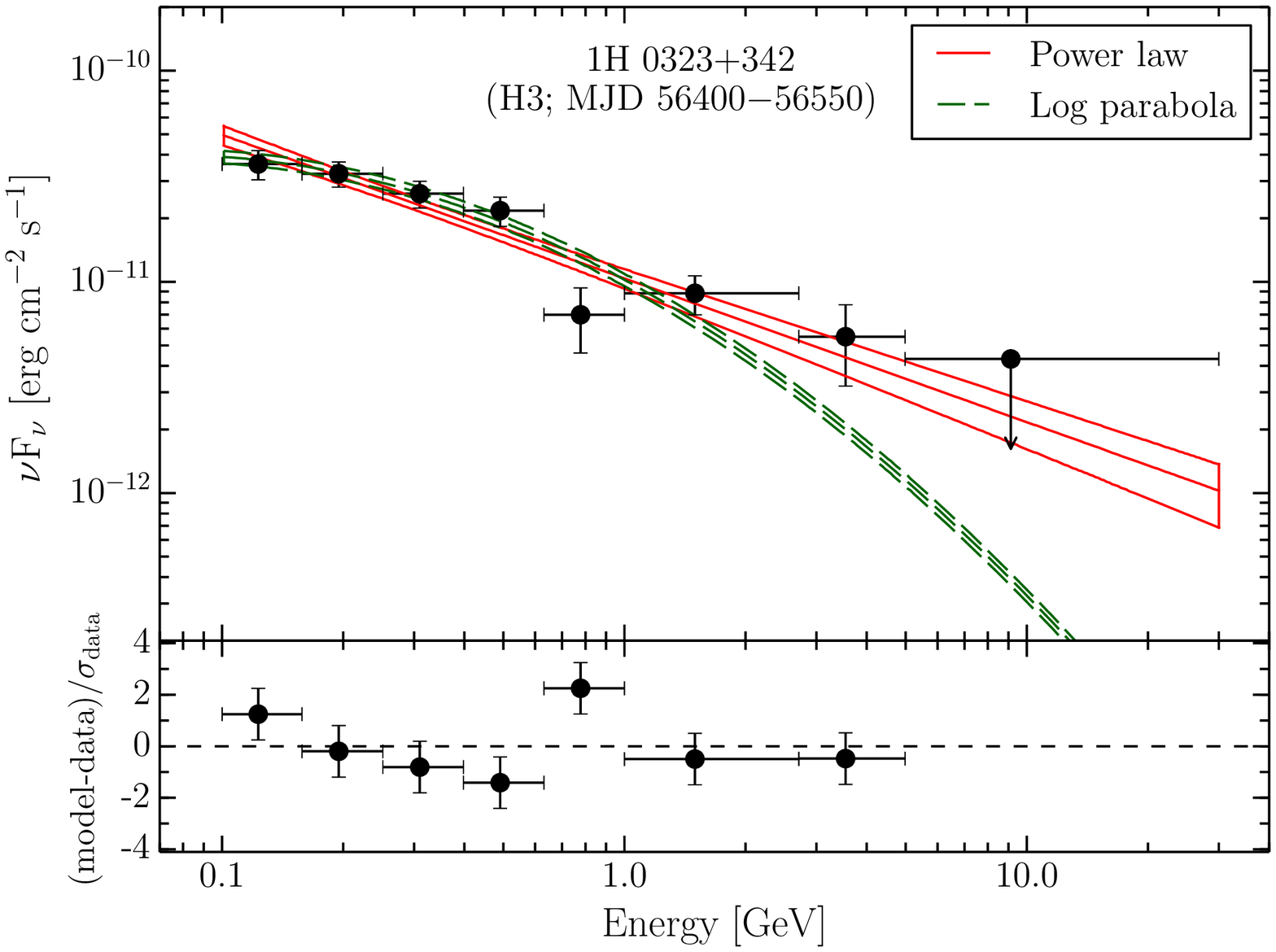}
\includegraphics[width=9.0cm,height=7.0cm]{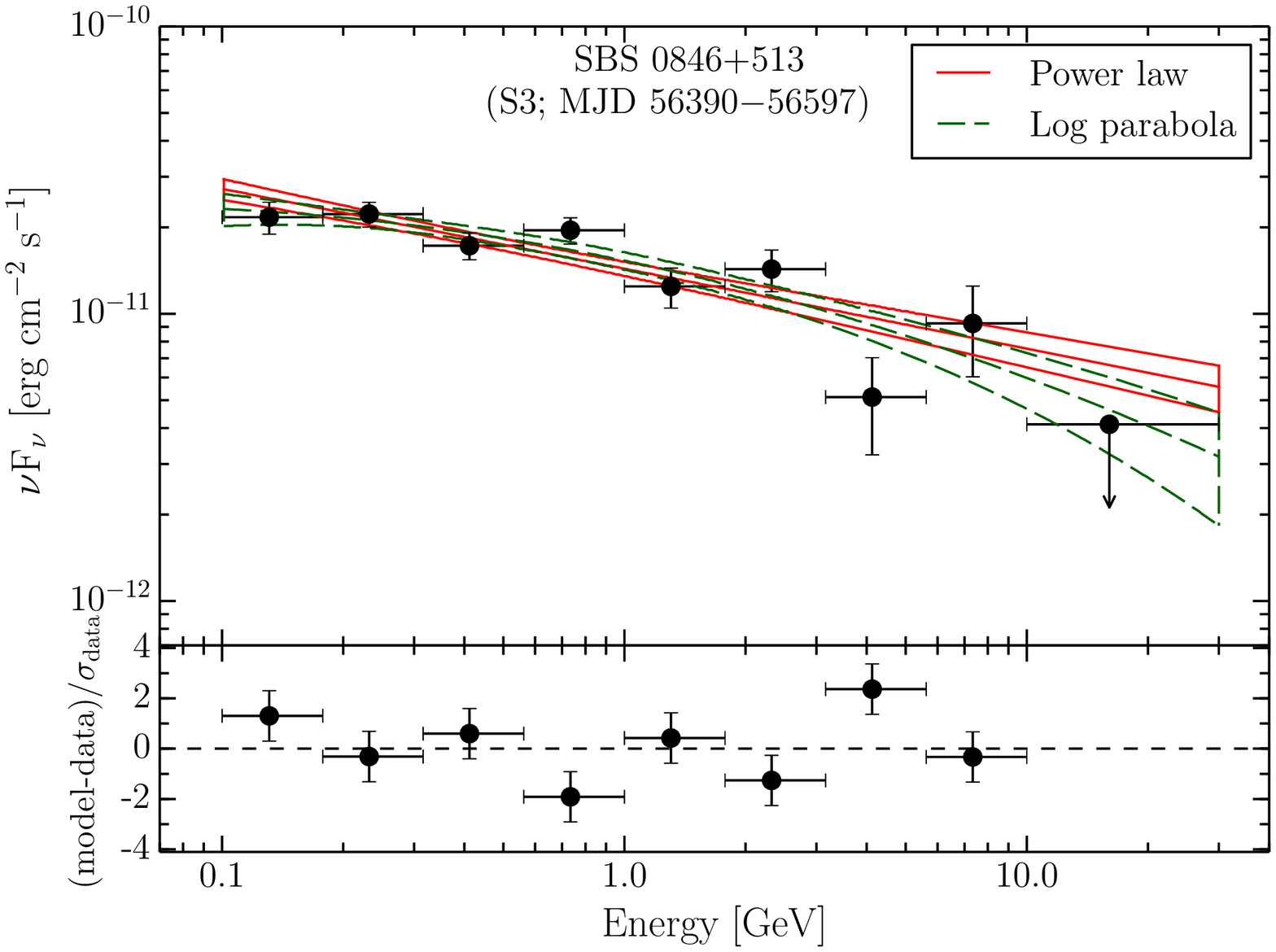}
}
\caption{{\it Fermi}-LAT SEDs of 1H 0323+342 (left panels) and SBS 0846+513 (right panels) in their different brightness states. Other information are same as in Figure~\ref{avg_spec_fig}.}\label{H3_P9}
\end{figure*}

\begin{figure*}
\hbox{\hspace{-0.5cm}
\includegraphics[width=9.0cm,height=7.0cm]{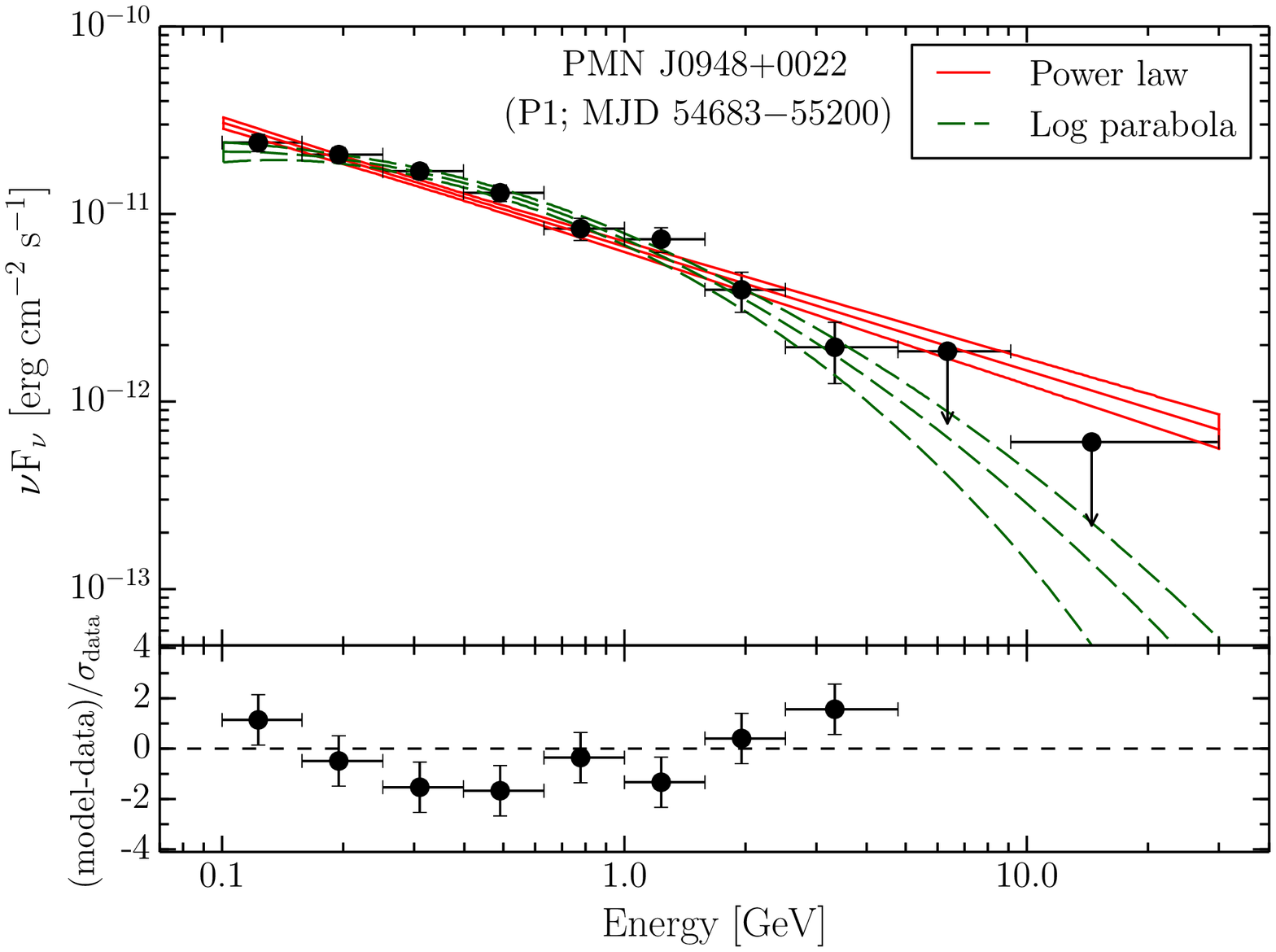}
\includegraphics[width=9.0cm,height=7.0cm]{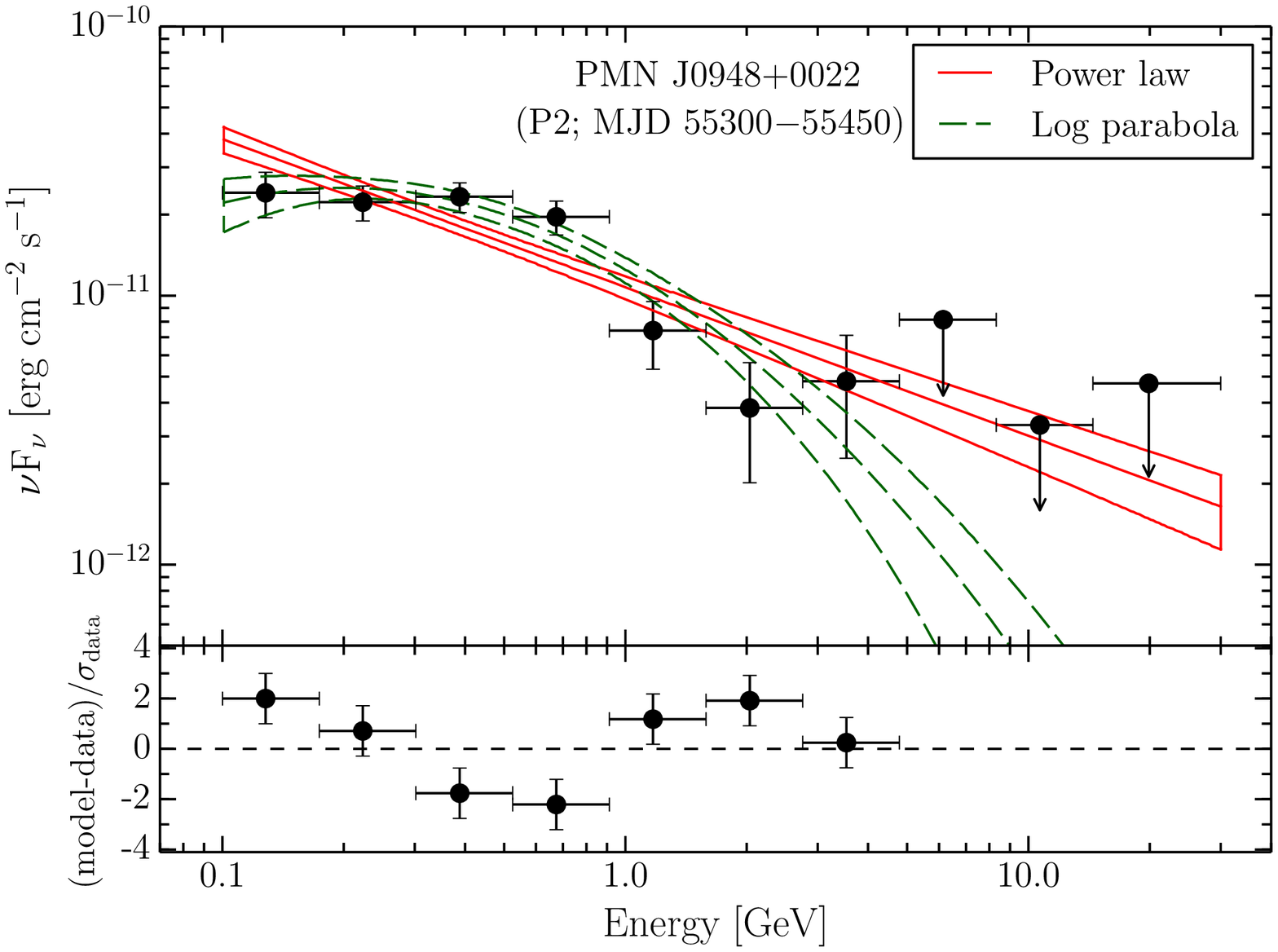}
}
\centering
\includegraphics[width=9.0cm,height=7.0cm]{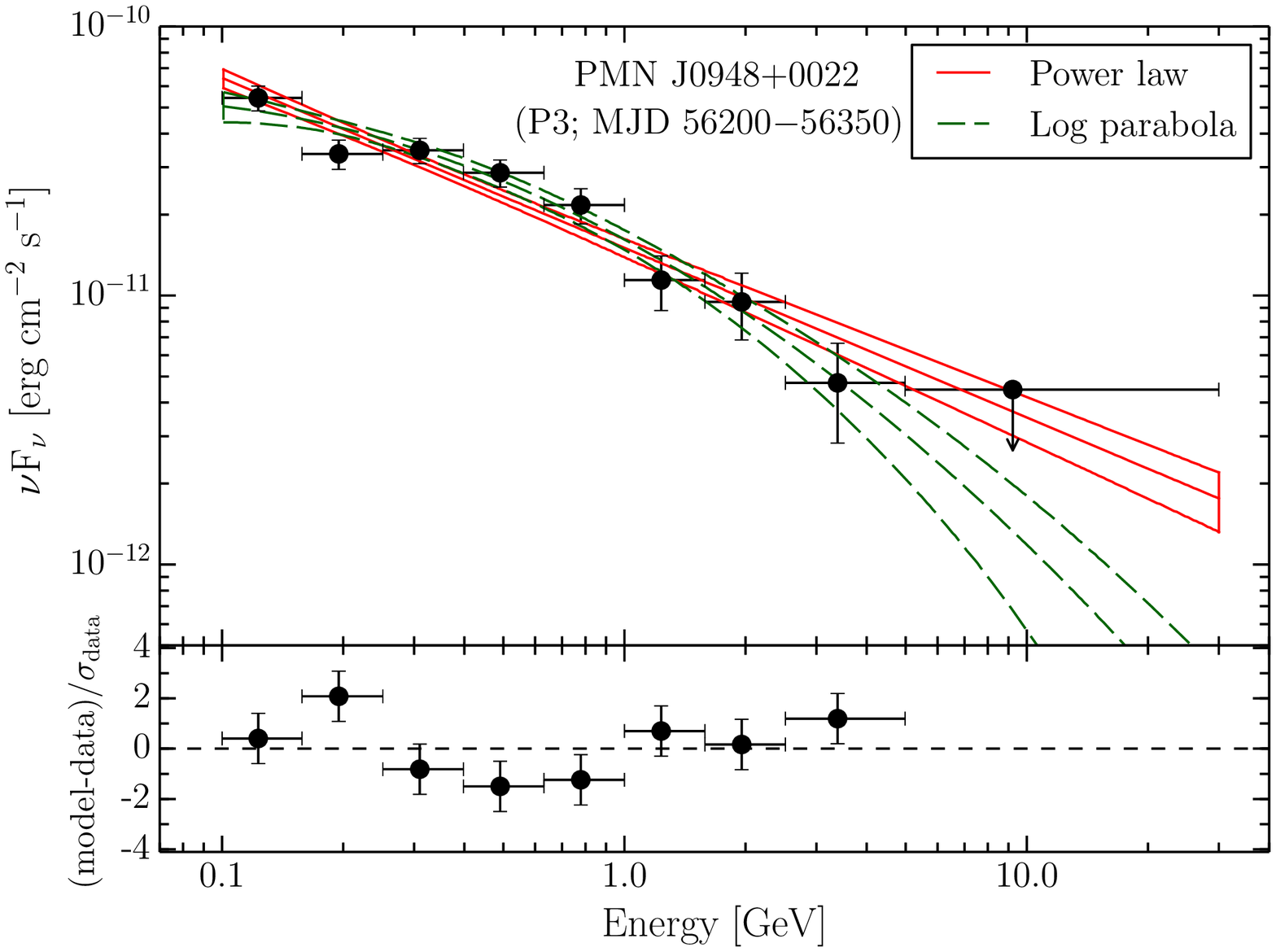}

\caption{{\it Fermi}-LAT SEDs of PMN J0948+0022 during its different activity states. Other information are same as in Figure~\ref{avg_spec_fig}.}\label{sbs}
\end{figure*}

Of the five $\gamma$-NLSy1 galaxies in our sample, three sources 1H 0323+342, SBS 0846+513, and PMN J0948+0022 are relatively bright and have shown GeV flaring activities \citep[e.g.,][]{2013ATel.5344....1C,2011ATel.3452....1D,2010ATel.2733....1D}, thus enabling us to probe their spectral behavior at different activity states. From the light curves, we identify periods when the sources are found to be in relatively low and high states. These periods are marked in Figure~\ref{lc}. A high activity state refers to the period of detection of high flux by the LAT, as already reported in the literature. A low activity state is selected as the time period when the source is in a relatively fainter state. We note that the low activity states may not represent the actual quiescence of the sources. The idea here is to compare the spectral shapes at different activity periods and selection of true quiescence would not serve the purpose due to poor photon statistics. Many GeV flares from PMN J0948+0022 have been reported \citep{2013ATel.4694....1D,2011ATel.3429....1D,2010ATel.2733....1D} and we select the two brightest flares (P2 and P3) for further analysis. The time duration along with the results of the likelihood analysis are presented in Table~\ref{flare_res}. For better visualization, we also show the spectra of the sources along with their best-fit model in Figure~\ref{H3_P9} and \ref{sbs}. A statistically significant ($\sim$ 4$\sigma$) curvature is noticed in the low activity state $\gamma$-ray spectrum of 1H 0323+342. Interestingly, no curvature is found in the high activity states of this source. A substantial curvature ($\sim 5\sigma$) is noticed in the first flaring state (S1) of SBS 0846+513 which is similar to that reported by \citet{2012MNRAS.426..317D}. However, no curvature is found in the other flaring state of this source (period S2; MJD 56,018$-$56,170), whereas for the same period \citet{2013MNRAS.436..191D} have reported the detection of curvature. A curvature is also noticed in the low activity state (P1) of PMN J0948+0022 but not during the GeV outbursts. This behavior is in contrast to that observed in bright blazars 3C 454.3 and PKS 1510$-$089 where a significant curvature is detected in their $\gamma$-ray spectra during the outbursts \citep{2011ApJ...733L..26A,2010ApJ...721.1425A}.

\subsubsection{Spectral Evolution}\label{spec_evol}
We investigate here the changes in the spectra in relation to the brightness of the $\gamma$-NLSy1 galaxies. Instead of attempting to provide a  best description of a spectrum, we intend to search for possible trends in it. Hence, we evaluate the PL spectral indices during the time bins used for the light curves shown in Figure~\ref{lc}. The spectral index can be considered as a representative of the mean slope.

The photon index of 1H 0323+342 is found to be harder in the high activity states. For example, in the bin centered at MJD 56,285 (peak flux in the interval H2 in Figure~\ref{lc}), the photon index value of 2.51~$\pm$~0.12 is significantly harder than the value of 2.78~$\pm$~0.05 found for the five year average. A hard photon index value of 2.49~$\pm$~0.11 is found during the 2013 August GeV outburst also. Similar behavior is also seen in SBS 0846+513. During its first GeV outburst, which led to the discovery of this source in the $\gamma$-ray band, the photon index is found to be 1.93$\pm$~0.09 which is typical of high synchrotron peaked blazars \citep{2011ApJ...743..171A}. Similar behavior of this source is earlier reported by \citet{2012MNRAS.426..317D}. The photon index of PMN J0948+0022 is also found to be harder during the high activity periods. During its first outburst of 2010 July, a photon index of 2.39~$\pm$~0.12 is obtained which is harder when compared to the five years averaged value of 2.62~$\pm$~0.02. This result is similar to that first reported by \citet{2011MNRAS.413.1671F}. A maximum flux of (1.21~$\pm$~0.36)~$\times$~10$^{-7}$~\phflux~is observed from PKS 1502+036 in the bin centered at MJD 55,477. The corresponding photon index during this period is 2.83~$\pm$~0.28 which is not significantly different from its five years averaged value of 2.63~$\pm$~0.05. Spectral evolution of PKS 2004$-$447 could not be ascertained owing to its faintness. Overall, though there are hints of spectral hardening for three out of the five $\gamma$-NLSy1 galaxies during flares, in comparison to their five year average behavior, no strong spectral change is noticed on shorter time scales. These results are in line with the findings of \citet{2010ApJ...710.1271A} and \citet{2012A&A...548A.106F}.

\begin{figure*}
\centering
\vbox
 {
\hbox{
      \includegraphics[width=9cm,height=7.0cm]{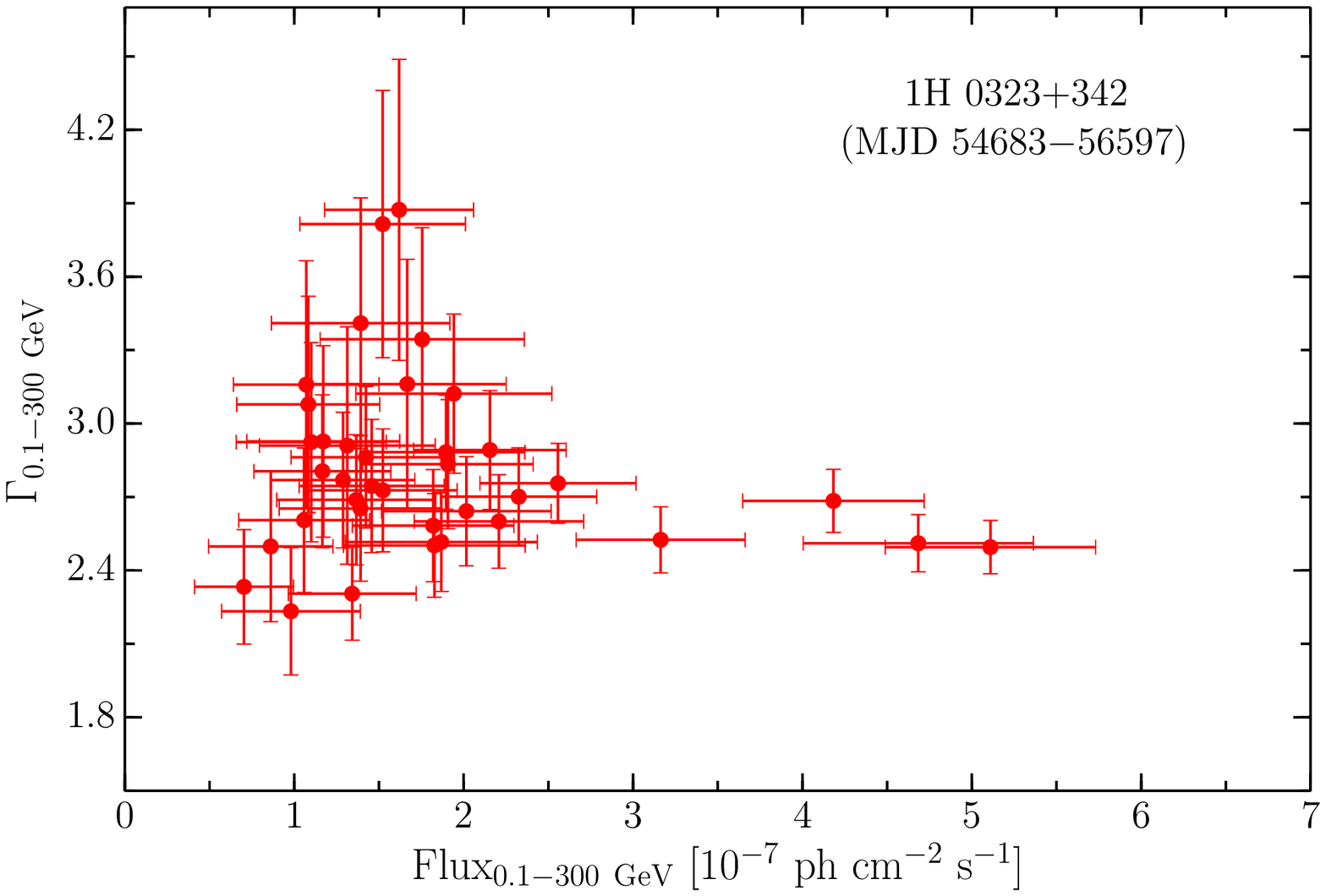}
      \includegraphics[width=9cm,height=7.0cm]{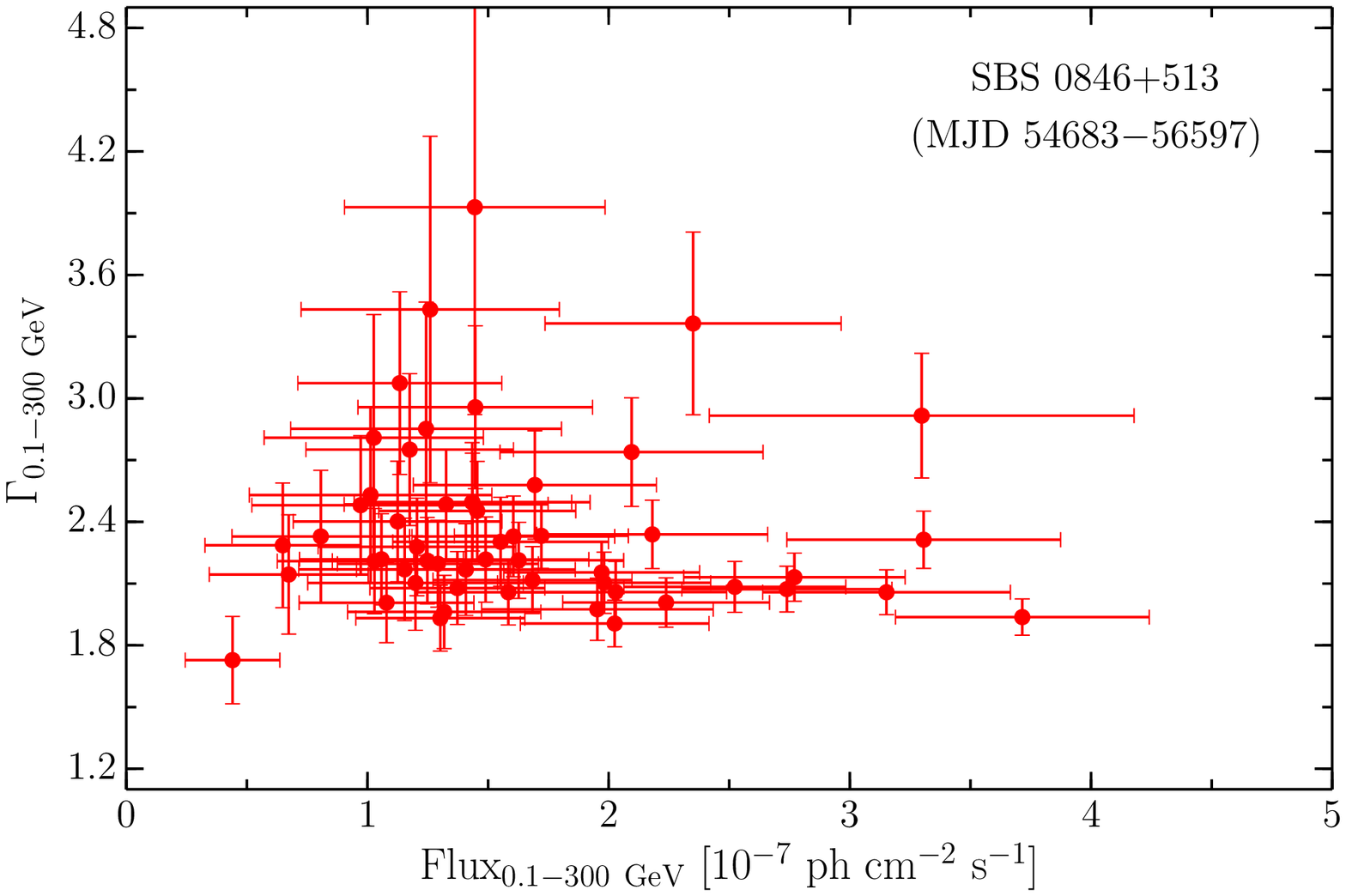}
     }
\hbox{
      \includegraphics[width=9cm,height=7.0cm]{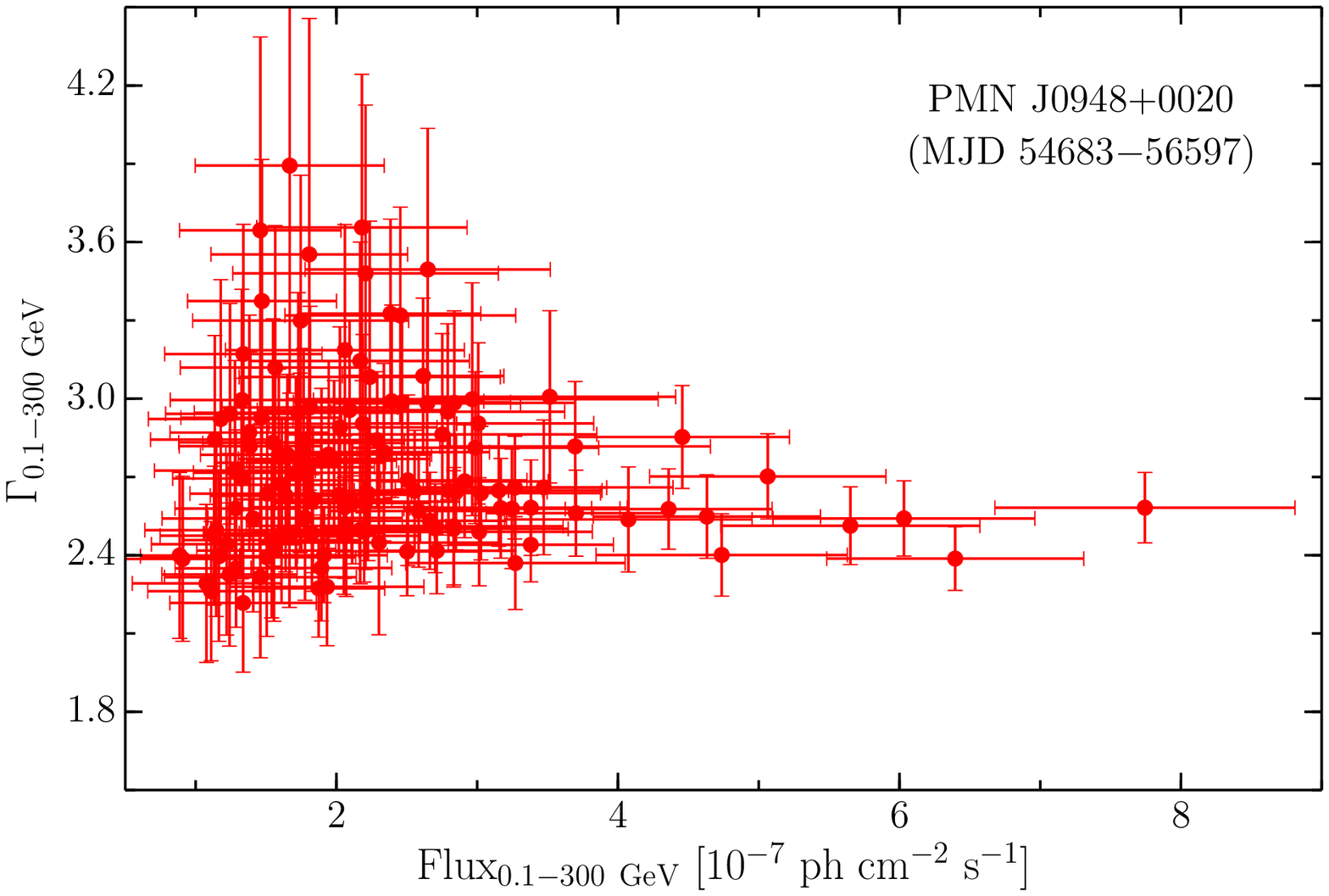}
      \includegraphics[width=9cm,height=7.0cm]{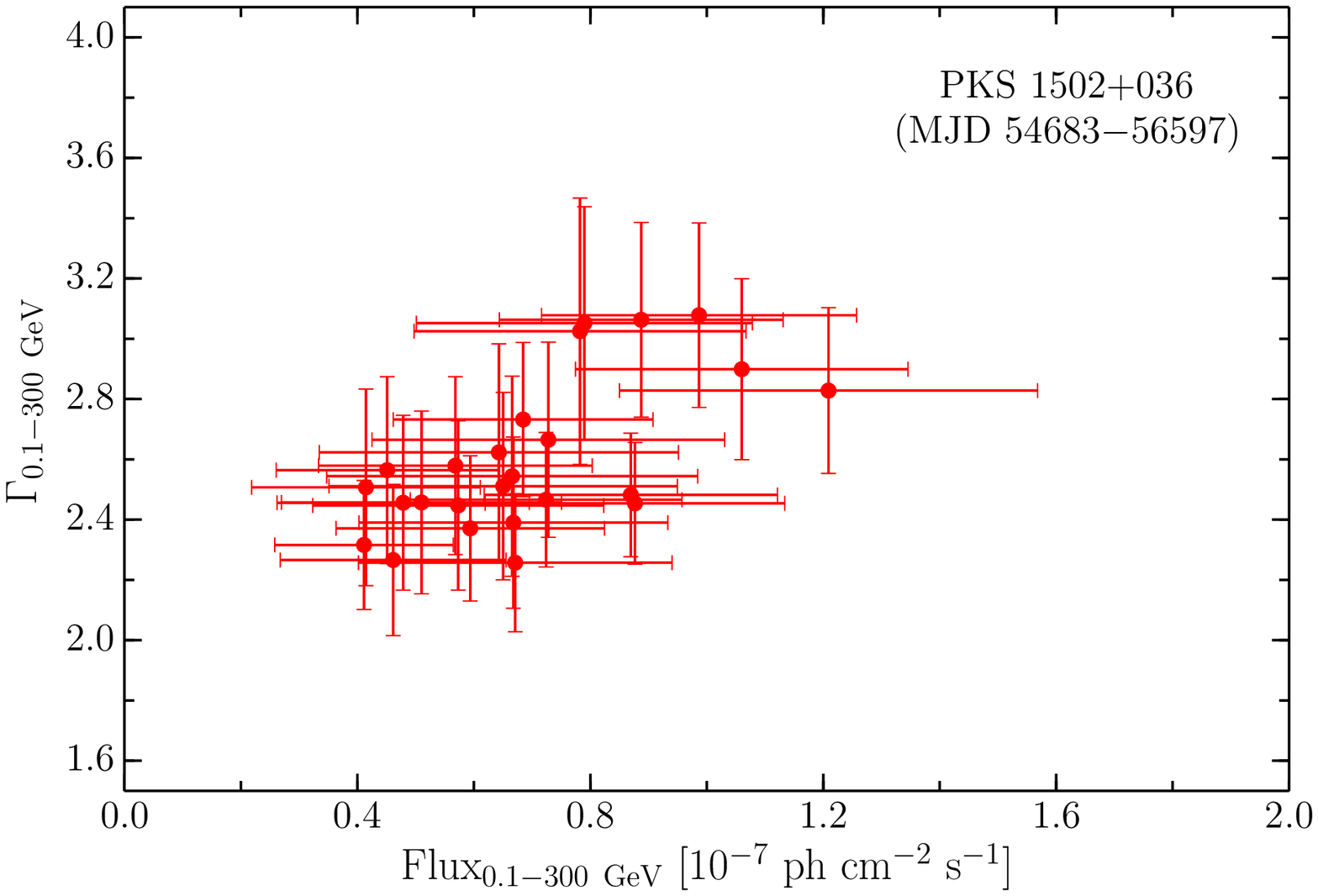}
     }
\includegraphics[width=9cm,height=7.0cm]{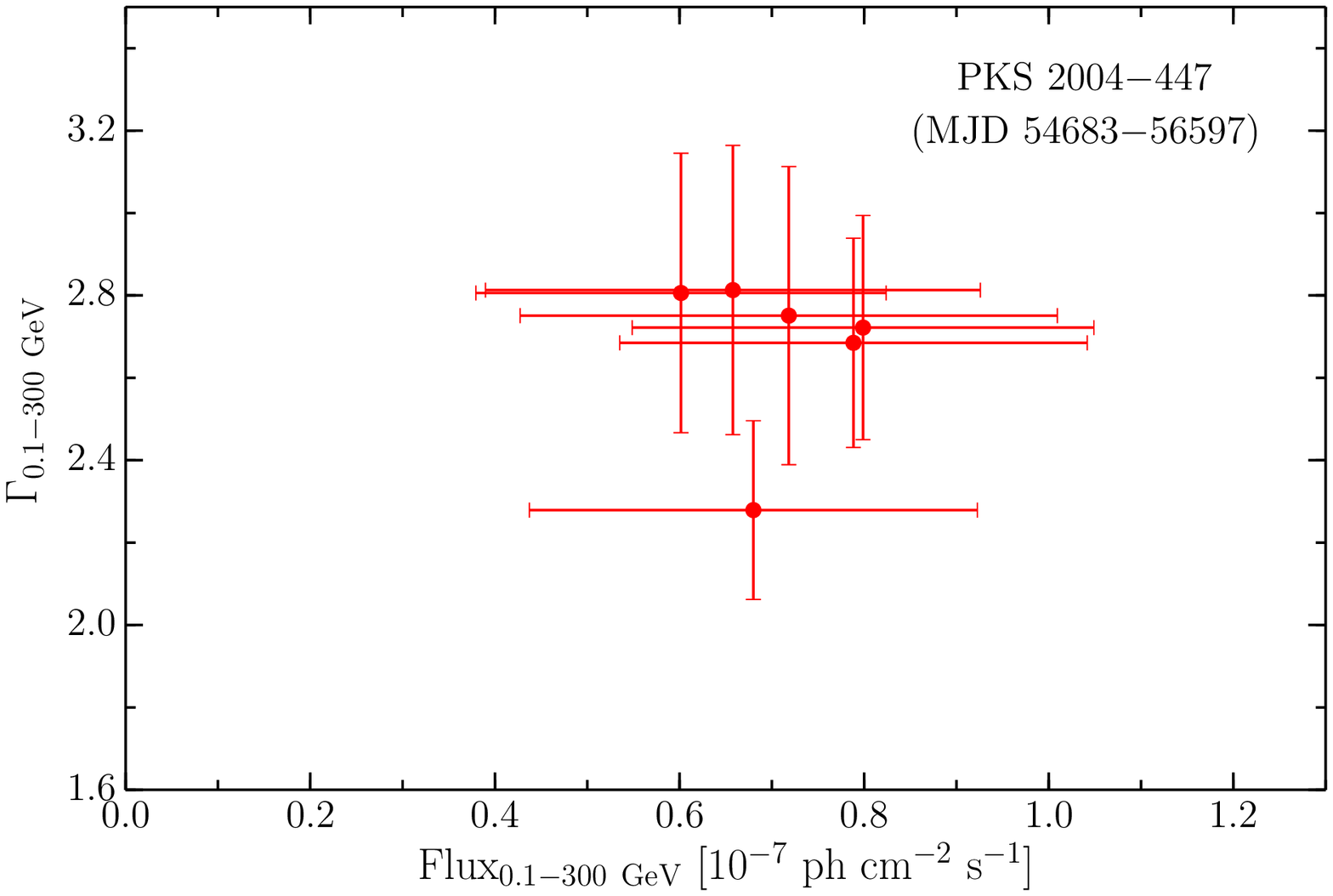}
  }
\caption{Photon index versus $\gamma$-ray flux plots, obtained from long term light curve analysis.}\label{ph_npred}
\end{figure*}

To study the overall spectral behavior of these sources, we plot in Figure~\ref{ph_npred} the photon index against the flux, obtained from the light curve analysis. PKS 1502+036 shows a `softer when brighter' behavior whereas no conclusion can be drawn about PKS 2004$-$447 owing to the small number of data points. Similar results have been reported by \citet{2013ApJ...768...52P} also. Moreover, the remaining three sources seem to show a `softer when brighter' trend, up to a flux level of $\simeq 1.5 \times 10^{-7}$ \phflux. Above this flux value, these sources possibly hint for a `harder when brighter' trend. To statistically test these behaviors, we perform a Monte Carlo test that takes into account of the dispersion in flux and photon index measurements. For each observed pair of flux and index values, we re-sample it by extracting the data from a normal distribution centered on the observed value and standard deviation equal to the 1$\sigma$ error estimate. We do not perform this test on the data of PKS 1502+036 and PKS 2004$-$447 due to their small sample size. For lower fluxes (i.e., $F_{\gamma} < 1.5 \times 10^{-7}$ \phflux), the correlation coefficients ($\rho$) are found to be 0.21, 0.26, and 0.27 with 95\% confidence limits of $-0.27 \leqslant \rho \leqslant 0.61, -0.03 \leqslant \rho \leqslant 0.51$, and $-0.05 \leqslant \rho \leqslant 0.54$ respectively for 1H 0323+342, SBS 0846+513, and PMN J0948+0022. On the other hand, for higher fluxes ($F_{\gamma} > 1.5 \times 10^{-7}$ \phflux), the correlation coefficients are $-$0.32, $-$0.09, and $-$0.08 with 95\% confidence limits of $-0.67 \leqslant \rho \leqslant 0.14, -0.48 \leqslant \rho \leqslant 0.33$, and $-0.27 \leqslant \rho \leqslant 0.12$ respectively for 1H 0323+342, SBS 0846+513, and PMN J0948+0022. Clearly, based on the Monte Carlo analysis, it is difficult to claim for the presence of correlation between fluxes and photon indices.

\subsection{Origin of Spectral Curvature/Break}\label{break_orign}
The origin of the $\gamma$-ray spectral break detected by {\it Fermi}-LAT in many FSRQs is still an open question. Many theoretical models are available in the literature to explain the observed spectral break. One among the extrinsic causes could be due to the attenuation of $\gamma$-rays by photon-photon pair production on He II Lyman recombination lines within the BLR as explained by the double-absorber model of \citet{2010ApJ...717L.118P}. In this model, the $\gamma$-ray dissipation region should lie inside the BLR. However, by analyzing a sample of $\gamma$-ray bright blazars, \citet{2012ApJ...761....2H} have raised questions about the validity of the double-absorber model. The break energy predicted by this model is found to be inconsistent with observations. \citet{2010ApJ...714L.303F} proposed a scenario wherein the observed spectral break could be explained as a sum of hybrid scattering of accretion disk and BLR photons, but this solution requires a wind like profile for the BLR. Using an equipartition approach, \citet{2013ApJ...771L...4C} proposed a scenario in which the break in the GeV spectra occurs as a consequence of the Klein-Nishina (KN) effect on the Compton scattering of BLR photons by relativistic electrons having a curved particle distribution in the jet. Alternatively, the $\gamma$-ray spectral break can also result from intrinsic effects. In this intrinsic model, the break can happen if there is a cutoff in the energy distribution of particles that produce the $\gamma$-ray emission \citep{2009ApJ...699..817A}.

Three out of the five $\gamma$-NLSy1 galaxies studied here, 1H 0323+342, SBS 0846+513 and PMN J0948+0022, show clear deviation from a simple PL model. Instead, the LP model seems to describe the data better. From the SED modeling of 1H 0323+342 and PMN J0948+0022, \citet{2009ApJ...707L.142A} constrained the $\gamma$-ray emission region in them to be well inside the BLR \citep[see also,][]{2012A&A...548A.106F,2014ApJ...789..143P}. In such a scenario, presence of the curvature in the $\gamma$-ray spectrum of these two sources can be explained on the basis of the KN effect. This is because the UV photons produced by the accretion disk and reprocessed by the BLR are blueshifted by a factor $\sim \Gamma$. Assuming a typical value of $\Gamma \simeq 10$, the approximate energy of these photons in the rest frame of the blob is about 10$^{16}$ Hz and hence the IC scattering with the electrons (having $\gamma \simeq 1000$) occurs under the mild KN regime. This hypothesis can also be tested by determining the energy of the highest energy photons detected by {\it Fermi}-LAT. The detection of very high energy photons ($>$ few tens of GeV) will clearly rule out the possibility of the emission region to be located inside the BLR. We used {\tt gtsrcprob} tool to determine the energy of the highest energy photon. This value is found to be 32.7 GeV (95.76\% detection probability) and 4.71 GeV (98.15\% detection probability) for 1H 0323+342 and PMN J0948+0022 respectively. The highest photon energy for 1H 0323+342 closely satisfies the BLR $\gamma$-ray transparency condition \citep{2009MNRAS.397..985G} and thus the origin of the $\gamma$-ray emission could lie inside the BLR. We note here that the non-detection of very high energy $\gamma$-ray photon does not ensure the emission region to be located inside the BLR but such a detection would have definitely ruled out this possibility.

The $\gamma$-ray emission from SBS 0846+513 is well modeled by IC scattering of dusty torus photons \citep{2013MNRAS.436..191D} and this sets the location of the $\gamma$-ray emission region to be outside the BLR where IC scattering takes place under the Thomson regime. The most probable cause of the curvature/break in the $\gamma$-ray spectrum would be, thus, intrinsic to the emitting particle distribution. In this picture, the curvature must be visible in the $\gamma$-ray spectrum of SBS 0846+513 in all brightness states. However, the poor photon statistics hinders us to detect this curvature. Additionally, if the observed curvature is due to the curved particle energy distribution, a similar feature should be seen in the optical/UV part of the spectrum also, provided that the optical/UV part is dominated by the synchrotron radiation. Hence, a dedicated multi-wavelength study, similar to the one done for 3C 454.3 \citep{2013ApJ...771L...4C}, is desired to test these possible scenarios.

\subsection{Gamma-ray loud NLSy1 Galaxies v/s Gamma-ray loud Blazars}\label{comparison}
The $\gamma$-ray spectral curvature observed in some of the $\gamma$-NLSy1 galaxies studied here are commonly known in {\it Fermi}-LAT detected FSRQs \citep{2010ApJ...710.1271A}. Such breaks are also noticed in the LAT spectra of a few low and intermediate synchrotron peaked BL Lac objects. In these BL lac objects where $\gamma$-ray spectral break is observed, broad optical emission lines are reported in the quiescent states of few of them. The sources AO 0235+164 \citep{2007A&A...464..871R} and PKS 0537$-$441 \citep{2011MNRAS.414.2674G} are good examples for this. It is thus likely that these BL Lac objects could belong to the FSRQ class of AGN as pointed by \citet{2011MNRAS.414.2674G}. Moreover, there are observations of curvature in the X-ray spectra of many high frequency peaked BL Lac objects \citep[and references therein]{2011ApJ...739...73M}, but the presence of curvature in their $\gamma$-ray spectra is yet to be verified. Therefore, based on the available observations it is likely that the $\gamma$-ray spectral break is a characteristic feature of the FSRQ class of AGN.

All the five sources studied here, have steep photon indices (Table~\ref{avg_res}) and the average value is found to be 2.55 $\pm$ 0.03. This is similar to the average photon index found for FSRQs (2.42 $\pm$ 0.17) in the 2LAC \citep{2011ApJ...743..171A} and is steeper than the value of 2.17 $\pm$ 0.12, 2.13 $\pm$ 0.14 and 1.90 $\pm$ 0.17 found for the low, intermediate and high synchrotron peaked BL Lac objects respectively. Thus, in terms of the average $\gamma$-ray photon index, these $\gamma$-NLSy1 galaxies are similar to FSRQs. However, the $\gamma$-ray luminosities of these five $\gamma$-NLSy1 galaxies are lower when compared to powerful FSRQs but higher than BL Lac objects \citep[e.g., Figure 6 of][]{2013ApJ...768...52P}. As some of the $\gamma$-NLSy1 galaxies studied here show evidences for the presence of spectral curvature and high flux variability, we argue based on the $\gamma$-ray properties, that these sources show resemblance to FSRQs.

Positioning the FSRQs and BL Lac objects detected during the first three months of {\it Fermi} operation \citep{2009ApJ...700..597A} on the $\gamma$-ray luminosity ($L_\gamma$) v/s $\gamma$-ray spectral index ($\alpha_{\gamma}$) diagram, \citet{2009MNRAS.396L.105G} have postulated the existence of low BH mass FSRQs having steep $\alpha_{\gamma}$ and low $L_\gamma$. In the $L_{\gamma}$ v/s $\alpha_{\gamma}$ plane, the
five $\gamma$-NLSy1 galaxies studied here tend to have steep $\alpha_{\gamma}$ values. Their $L_{\gamma}$ values are intermediate to the FSRQ 3C 454.3 and the BL Lac object Mrk 421 \citep{2013ApJ...768...52P}. Further, from the estimates of BH masses available in the literature, it is also found that these $\gamma$-NLSy1 galaxies host low mass BHs (Table~\ref{general}). They appear to be on the higher side of the BH masses known for NLSy1 galaxies in general \citep{2006AJ....132..531K,2008ApJ...685..801Y}, but lower than that known for powerful
blazars \citep{2010MNRAS.402..497G}. It is thus likely that these $\gamma$-NLSy1 galaxies are the low BH mass FSRQs already predicted by \citet{2009MNRAS.396L.105G}. However, we note that the estimation of the BH mass of NLSy1 galaxies is still a matter of debate. For example, using the virial relationship \citet{2008ApJ...685..801Y} reported the central BH mass of PMN J0948+0022 as $\sim$~10$^{7.5}~M_{\odot}$, whereas using the SED modeling approach \citet{2009ApJ...707L.142A} found $\sim$~10$^{8.2}~M_{\odot}$. Recently \citet{2013MNRAS.431..210C} obtained the BH mass of this source as large as $\sim$~10$^{9.2}~M_{\odot}$. Further, on examination of the SEDs of these $\gamma$-NLSy1 galaxies, 1H 0323+342 \citep{2014ApJ...789..143P}, SBS 0846+513 \citep{2012MNRAS.426..317D}, PMN J0948+0022 \citep{2011MNRAS.413.1671F}, PKS 1502+036 and PKS 2004$-$447 \citep{2013ApJ...768...52P}, we find all of them to have Compton dominance (which is the ratio of inverse Compton to synchrotron peak luminosities in the SED) of the order of $\sim$ 10 similar to that known for powerful FSRQs.

\section{Conclusions}\label{conclusion}
In this paper, we present the detailed analysis of the $\gamma$-ray temporal and spectral behavior of the five RL-NLSy1 galaxies that are detected by {\it Fermi}-LAT with high significance. We summarize our main results as follows:
\begin{enumerate}
\item During the period 2008 August to 2013 November, three out of five $\gamma$-NLSy1 galaxies show significant flux variations. The same cannot be claimed for the remaining two sources due to their faintness and/or intrinsic low variability nature. The $F_{\rm var}$ values obtained are 0.498 $\pm$ 0.046, 0.467 $\pm$ 0.042 and 0.443 $\pm$ 0.029 for 1H 0323+342, SBS 0846+513, and PMN J0948+0022 respectively. These are larger than that reported for FSRQs ($0.24 \pm 0.01$) as well as BL Lac objects ($0.12 \pm 0.01, 0.07 \pm 0.02$, and $0.07 \pm 0.02$ for LSP, ISP, and HSP BL Lac objects respectively) by \citet{2010ApJ...722..520A}. This supports the idea that in terms of the $\gamma$-ray flux variations these sources are similar to the FSRQ class of AGN. More than one episode of flaring activities is seen in 1H 0323+342, SBS 0846+513 and PMN J0948+0022, whereas no such flaring activities is observed in PKS 1502+342 and PKS 2004$-$447. Results from Monte Carlo simulation, to test for the presence of a possible correlation between the fluxes and photon indices, do not support for any correlation both at lower as well as at higher flux levels.

\item The observed average MeV$-$GeV $\gamma$-ray energy spectra of 1H 0323+342 and PMN J0948+0022 show significant deviation from a simple PL model. The data are well fit with the LP model. A hint for the presence of a curvature is found for SBS 0846+513. For PKS 1502+036 and PKS 2004$-$447, the PL model fits the observations very well.

\item Three sources, 1H 0323+342, SBS 0846+513 and PMN J0948+0022 have shown flux variations in their $\gamma$-ray emission, thereby enabling us to study their spectral behavior in different brightness states. Again PL model do not well represent the observed spectra in some of their activity states. A statistically significant curvature is observed in the low, high and moderately active state $\gamma$-ray spectra of 1H 0323+342, SBS 0846+513 and PMN J0948+0022 respectively.

\item The SED modeling of 1H 0323+342 and PMN J0948+0022 by \citet{2009ApJ...707L.142A} leads to the conclusion that the location of $\gamma$-ray emission in these sources is inside the BLR \citep[see also,][]{2012A&A...548A.106F,2014ApJ...789..143P}. It is thus likely that the
observed $\gamma$-ray emission in these sources is due to IC scattering of BLR photons occurring under the KN regime. In such a scenario, a prominent curvature in the $\gamma$-ray spectra is expected \citep[see][for a similar argument]{2010ApJ...721.1425A}. For SBS 0846+513, the $\gamma$-ray emission is well modeled by IC scattering of dusty torus photons \citep{2013MNRAS.436..191D}. Since the production of $\gamma$-rays would be in the Thomson regime, the observed curvature could be intrinsic to the source and can be attributed to the break/curvature in the particle energy distribution.

\item The average photon indices of the $\gamma$-NLSy1 galaxies, obtained from the PL fits, have values ranging between 2.2 and 2.8. This matches well with the average photon index (2.42 $\pm$ 0.17) found for FSRQs in the 2LAC. However these values are softer when compared to BL Lac objects \citep{2011ApJ...743..171A}.

With the accumulation of more data by {\it Fermi} in the future and subsequent generation of high quality $\gamma$-ray spectra, it will be possible to constrain further various models proposed for the origin of MeV$-$GeV break and other $\gamma$-ray spectral properties of the $\gamma$-NLSy1 galaxies. Finally, as observed by \citet{2010ApJ...710.1271A} and \citet{2011ApJ...733...19T}, a break in the $\gamma$-ray spectrum is a characteristic feature of powerful FSRQs, which is also very clearly visible in the case of $\gamma$-NLSy1 galaxies. The origin of such curvature and break in their spectra could be because of different physical mechanisms taking place at different regions and at different scales. The overall observed properties of these $\gamma$-NLSy1 galaxies, indicate that they could be similar to powerful FSRQs but with low or moderate jet power \citep[see][for a comparison]{2009ApJ...707L.142A,2013MNRAS.436..191D,2013ApJ...768...52P}.
\end{enumerate}

\section*{Acknowledgments}

We thank the referee for constructive comments that improved the presentation significantly. Use of {\it Hydra} cluster at the Indian Institute of Astrophysics is acknowledged.

\bibliographystyle{apj}
\bibliography{Master}

\newpage
\begin{table*}
\caption{Basic information of the $\gamma$-NLSy1 galaxies studied in this work.} 
\begin{center}
\begin{tabular}{lccccrrcc}
\hline \hline
Name  & RA (2000) & Dec (2000) & $z$      & V     & $\alpha_R$ & R & Log $M_{\rm BH}$ & References for $M_{\rm BH}$\\
      &  (h m s)  & (d m s)    &        & (mag)   &            &   &   (\Msun)      &                      \\
\hline
1H 0323+342    & 03:24:41.2 & +34:10:45 & 0.063 & 15.72 & -0.35 & 318 & 7.0 & \citet{2006ApJS..166..128Z} \\
SBS 0846+513   & 08:49:58.0 & +51:08:28 & 0.584 & 18.78 & -0.26& 4496 & 7.4 & \citet{2008ApJ...685..801Y} \\
PMN J0948+0022 & 09:48:57.3 & +00:22:24 & 0.584 & 18.64 & 0.81 & 846  & 7.5 & \citet{2008ApJ...685..801Y}\\
PKS 1502+036   & 15:05:06.5 & +03:26:31 & 0.409 & 18.64 & 0.41 & 3364 & 6.6 & \citet{2008ApJ...685..801Y} \\
PKS 2004$-$447 & 20:07:55.1 & -44:34:43 & 0.240 & 19.30 &0.38  & 6358 & 6.7 & \citet{2001ApJ...558..578O} \\
\hline
\end{tabular}
\end{center}
\label{general}
\end{table*}

\begin{table*}
\caption{Fractional rms variability and variability probability of the $\gamma$-NLSy1 galaxies.} 
\begin{center}
\begin{tabular}{lccc}
\hline \hline
Name  & Binning & $F_{\rm var}$ & Probability\\
      &  (days)  &  &  (\%)      \\
\hline
1H 0323+342    & 14 & 0.498~$\pm$~0.046 & $>$~99.0 \\
SBS 0846+513   & 7 & 0.467~$\pm$~0.042 & $>$~99.0  \\
PMN J0948+0022 & 7 & 0.443~$\pm$~0.029 & $>$~99.0 \\
PKS 1502+036   & 30 & -- & 34.5 \\
PKS 2004$-$447 & 30 & --& -- \\
\hline
\end{tabular}
\end{center}
\label{statistic}
\end{table*}

\begin{table}
\caption{Details of the PL and LP model fits to the five years averaged $\gamma$-ray data of the $\gamma$-NLSy1 galaxies.}.
\begin{flushleft}
 
\begin{tabular}{lcccccccccc}
\hline \hline

  &  \multicolumn{4}{c}{PL}     &   \multicolumn{5}{c}{LP} \\
    \cline{2-5}                & \cline{5-9} & \\
Name & $\Gamma_{\gamma}$ & $F_{0.1-300~{\rm GeV}}$\tablenotemark{a} &  TS &log $L_{\rm PL}$\tablenotemark{b} & $\alpha$ & $\beta$  &$F_{0.1-300~{\rm GeV}}$\tablenotemark{a}  & TS & log $L_{\rm LP}$\tablenotemark{b} & $TS_{\rm curve}$ \\

\hline
1H 0323+342     & 2.78 $\pm$ 0.05 & 7.54 $\pm$ 0.39  & 732.3  & 44.4 & 2.66 $\pm$ 0.06 & 0.19 $\pm$ 0.05 & 6.98 $\pm$ 0.40  & 738.3 & 44.3 & 17.36\\
SBS 0846+513    & 2.26 $\pm$ 0.03 & 4.35 $\pm$ 0.22  & 1666.9 & 46.7 & 2.04 $\pm$ 0.07 & 0.09 $\pm$ 0.03 & 3.73 $\pm$ 0.27  & 1660.4& 46.7 & 15.03\\
PMN J0948+0022  & 2.62 $\pm$ 0.02 & 13.03 $\pm$ 0.36 & 3299.8 & 47.0 & 2.35 $\pm$ 0.04 & 0.20 $\pm$ 0.03 & 11.78 $\pm$ 0.37 & 3386.7& 46.9 & 76.51\\
PKS 1502+036    & 2.63 $\pm$ 0.05 & 4.53 $\pm$ 0.35  & 434.8  & 46.1 & 2.53 $\pm$ 0.10 & 0.06 $\pm$ 0.05 & 4.29 $\pm$ 0.39  & 434.4 & 46.0 & 1.90 \\
PKS 2004$-$447  & 2.38 $\pm$ 0.09 & 1.42 $\pm$ 0.24  & 106.1  & 45.2 & 1.54 $\pm$ 0.36 & 0.42 $\pm$ 0.18 & 0.91 $\pm$ 0.23  & 113.1 & 45.3 & 10.10 \\
\hline
\end{tabular}
\tablenotetext{1}{The $\gamma$-ray flux values in units of 10$^{-8}$ \phflux}
\tablenotetext{2}{The PL and LP luminosities at logarithmic scale}
\label{avg_res}
\\
\end{flushleft}
\end{table}

\begin{table}
\caption{Details of the PL and LP model fits to the various brightness states of the $\gamma$-NLSy1 galaxies. The units are same as in Table~\ref{avg_res}.}
\begin{flushleft}
 
\begin{tabular}{lcccccccccc}
\hline \hline

  &  \multicolumn{4}{c}{PL}     &   \multicolumn{5}{c}{LP} \\
    \cline{2-5}                & \cline{5-9} & \\
Activity state (MJD) & $\Gamma_{\gamma}$ & $F_{0.1-300~{\rm GeV}}$ &  TS & L$_{\rm PL}$ & $\alpha$ & $\beta$  &$F_{0.1-300~{\rm GeV}}$  & TS  &L$_{\rm LP}$ & $TS_{\rm curve}$ \\

\hline
     &  &   &  & 1H 0323+342 &  & &  &  & \\
\\
H1 (54683$-$56200)     & 2.83 $\pm$ 0.07 & 5.44 $\pm$ 0.46   & 291.8 & 44.3 & 2.70 $\pm$ 0.09 & 0.34 $\pm$ 0.11 & 4.89 $\pm$ 0.43   & 301.7 & 44.2 & 16.13\\
H2 (56200$-$56400)     & 2.67 $\pm$ 0.08 & 12.90 $\pm$ 1.23  & 222.0 & 44.7 & 2.49 $\pm$ 0.12 & 0.21 $\pm$ 0.09 & 11.60 $\pm$ 1.28  & 228.0 & 44.6 & 7.68\\
H3 (56400$-$56550)     & 2.68 $\pm$ 0.07 & 18.50 $\pm$ 1.51  & 335.2 & 44.9 & 2.62 $\pm$ 0.09 & 0.08 $\pm$ 0.06 & 17.80 $\pm$ 1.56  & 335.0 & 44.7 & 1.61\\
\hline
     &  &   &  & SBS 0846+513 &  &&  &  & \\
\\
S1 (55716$-$55746)     & 1.99 $\pm$ 0.01 & 21.90 $\pm$ 0.52  & 608.6  & 47.6 & 1.21 $\pm$ 0.01 & 0.32 $\pm$ 0.01 & 15.60 $\pm$ 0.22 & 634.8 & 47.8 & 29.53\\
S2 (56018$-$56170)     & 2.14 $\pm$ 0.05 & 12.20 $\pm$ 0.90  & 917.0  & 47.2 & 1.92 $\pm$ 0.09 & 0.09 $\pm$ 0.03 & 10.80 $\pm$ 0.95 & 919.3 & 47.2 & 6.87 \\
S3 (56390$-$56597)     & 2.24 $\pm$ 0.05 & 12.20 $\pm$ 0.81  & 983.9  & 47.1 & 2.08 $\pm$ 0.08 & 0.07 $\pm$ 0.03 & 11.20 $\pm$ 0.87 & 981.5 & 47.1 & 5.20 \\
\hline
     &  &   &  & PMN J0948+0022 &&  &  &  & \\
\\
P1 (54683$-$55200)     & 2.67 $\pm$ 0.05 & 11.70 $\pm$ 0.63  & 842.9 & 46.9 &  2.40 $\pm$ 0.08 & 0.21 $\pm$ 0.05 & 10.70 $\pm$ 0.62  & 866.0 & 46.9 & 21.72\\
P2 (55300$-$55450)     & 2.54 $\pm$ 0.06 & 14.40 $\pm$ 1.22  & 323.2 & 47.1 &  2.14 $\pm$ 0.17 & 0.26 $\pm$ 0.06 & 12.20 $\pm$ 2.13  & 331.2 & 47.0 & 12.02 \\
P3 (56200$-$56350)     & 2.62 $\pm$ 0.01 & 24.00 $\pm$ 0.47  & 756.9 & 47.3 &  2.44 $\pm$ 0.01 & 0.14 $\pm$ 0.01 & 22.60 $\pm$ 0.19  & 764.5 & 47.2 & 7.65\\
\hline
\end{tabular}
\label{flare_res}
\\
\end{flushleft}
\end{table}

\newpage

\newpage

\end{document}